\documentclass[]{aastex631}

\usepackage{amsmath}

\shorttitle{GRB\,200613A}
\shortauthors{Fu et al.}

\graphicspath{{./}{figures/}}

\begin{document}

\title{Unveiling the Multifaceted GRB 200613A: Prompt Emission Dynamics, Afterglow Evolution, and the Host Galaxy's Properties}

\correspondingauthor{Dong Xu, Wei-Hua Lei}
\email{dxu@nao.cas.cn, leiwh@hust.edu.cn}

\author[0009-0002-7730-3985]{Shao-Yu Fu}
\affiliation{Key Laboratory of Space Astronomy and Technology, National Astronomical Observatories, Chinese Academy of Sciences, Beijing, 100101, China}
\affiliation{School of Astronomy and Space Science, University of Chinese Academy of Sciences, Chinese Academy of Sciences, Beijing 100049, China}

\author[0000-0003-3257-9435]{Dong Xu}
\affiliation{Key Laboratory of Space Astronomy and Technology, National Astronomical Observatories, Chinese Academy of Sciences, Beijing, 100101, China}

\author[0000-0003-3440-1526]{Wei-Hua Lei}
\affiliation{Department of Astronomy, School of Physics, Huazhong University of Science and Technology, Wuhan, 430074, China}

\author[0000-0001-7717-5085]{Antonio~de~Ugarte~Postigo}
\affiliation{Universit\'{e} de la C\^ote d'Azur, Observatoire de la C\^ote d'Azur, CNRS, Artemis, 06304 Nice, France}
\affiliation{Aix Marseille Univ, CNRS, LAM Marseille, France}

\author[0000-0003-2902-3583]{D. Alexander Kann}
\affiliation{Hessian Research Cluster ELEMENTS, Giersch Science Center, Max-von-Laue-Straße 12, Goethe University Frankfurt, Campus Riedberg, D-60438 Frankfurt am Main, Germany}

\author[0000-0002-7978-7648]{Christina~C.~Th\"one}
\affiliation{Astronomical Institute, Czech Academy of Sciences, Fri\v cova 298, Ond\v rejov, Czech Republic}

\author[0000-0001-6991-7616]{Jos\'e~Feliciano~Ag\"u\'i~Fern\'andez}
\affiliation{Centro Astron\'omico Hispano en Andaluc\'ia, Observatorio de Calar Alto, Sierra de los Filabres, 04550 G\'ergal, Almer\'ia, Spain}

\author{Yi Shuang-Xi}
\affiliation{School of Physics and Physical Engineering, Qufu Normal University, Qufu 273165, China}

\author[0000-0001-5553-4577]{Wei Xie}
\affiliation{Department of Astronomy, School of Physics and Electronic Science, Guizhou Normal University, Guiyang 550001, China}
\affiliation{Guizhou Provincial Key Laboratory of Radio Astronomy Data Processing, Guizhou Normal University, Guiyang 550001, China}

\author[0000-0002-5400-3261]{Yuan-Chuan Zou}
\affiliation{Department of Astronomy, School of Physics, Huazhong University of Science and Technology, Wuhan, 430074, China}

\author{Xing Liu}
\author{Shuai-Qing Jiang}
\author{Tian-Hua Lu}
\author{Jie An}
\affiliation{Key Laboratory of Space Astronomy and Technology, National Astronomical Observatories, Chinese Academy of Sciences, Beijing, 100101, China}
\affiliation{School of Astronomy and Space Science, University of Chinese Academy of Sciences, Chinese Academy of Sciences, Beijing 100049, China}

\author[0000-0002-9022-1928]{Zi-Pei Zhu}
\affiliation{Key Laboratory of Space Astronomy and Technology, National Astronomical Observatories, Chinese Academy of Sciences, Beijing, 100101, China}

\author[0000-0001-6637-6973]{Jie Zheng}
\affiliation{CAS Key Laboratory of Optical Astronomy, National Astronomical Observatories, Chinese Academy of Sciences, Beijing 100101, China}

\author{Qing-Wen Tang}
\author{Peng-Wei Zhao}
\affiliation{Department of Physics, School of Science, Nanchang University, Nanchang 330031, China}

\author[0000-0002-9422-3437]{Li-Ping Xin}
\affiliation{Key Laboratory of Space Astronomy and Technology, National Astronomical Observatories, Chinese Academy of Sciences, Beijing, 100101, China}

\author{Jian-Yan Wei}
\affiliation{Key Laboratory of Space Astronomy and Technology, National Astronomical Observatories, Chinese Academy of Sciences, Beijing, 100101, China}
\affiliation{School of Astronomy and Space Science, University of Chinese Academy of Sciences, Chinese Academy of Sciences, Beijing 100049, China}

\begin{abstract}

We present our optical observations and multi-wavelength analysis of the GRB\,200613A detected by \texttt{Fermi} satellite. Time-resolved spectral analysis of the prompt $\gamma$-ray emission was conducted utilizing the Bayesian block method to determine statistically optimal time bins. Based on the Bayesian Information Criterion (BIC), the data generally favor the Band+Blackbody (short as BB) model. We speculate that the main Band component comes from the Blandford-Znajek mechanism, while the additional BB component comes from the neutrino annihilation process. The BB component becomes significant for a low-spin, high-accretion rate black hole central engine, as evidenced by our model comparison with the data. The afterglow light curve exhibits typical power-law decay, and its behavior can be explained by the collision between the ejecta and constant interstellar medium (ISM). Model fitting yields the following parameters: $E_{K,iso} = (2.04^{+11.8}_{-1.50})\times 10^{53}$ erg, $\Gamma_0=354^{+578}_{-217}$, $p=2.09^{+0.02}_{-0.03}$, $n_{18}=(2.04^{+9.71}_{-1.87})\times 10^{2}$ cm$^{-3}$, $\theta_j=24.0^{+6.50}_{-5.54}$ degree, $\epsilon_e=1.66^{+4.09}_{-1.39})\times 10^{-1}$ and $\epsilon_B=(7.76^{+48.5}_{-5.9})\times 10^{-6}$. In addition, we employed the public Python package \texttt{Prospector} perform a spectral energy distribution (SED) modeling of the host galaxy. The results suggest that the host galaxy is a massive galaxy ($\log(M_\ast / M_\odot)=11.75^{+0.10}_{-0.09}$) with moderate star formation rate ($\mbox{SFR}=22.58^{+13.63}_{-7.22} M_{\odot}$/yr). This SFR is consistent with the SFR of $\sim 34.2 M_{\odot}$ yr$^{-1}$ derived from the [OII] emission line in the observed spectrum.

\end{abstract}

\keywords{\href{http://astrothesaurus.org/uat/629}{Gamma-ray bursts (629)}}

\section{Introduction} \label{sec:intro}
Gamma-ray bursts (GRBs) are the most energetic and luminous explosions in the universe, which release isotropic energy $10^{48}-10^{55}$ ergs within seconds \citep{2015PhR...561....1K}.  Their exceptional luminosity enables their detection even at remarkably distant locations, reaching redshifts up to $z\sim 9$ \citep{2011ApJ...736....7C}. The study of GRBs has provided invaluable insights into the physics of extreme astrophysical phenomena and the evolution of the early Universe \citep{2006AIPC..836..503B}. 

GRBs can be categorized into two primary types based on their duration: long and short \citep{1993ApJ...413L.101K}.  Long GRBs, which constitute about 70\% of all GRBs, typically last for a few seconds to a few minutes and exhibit a high energy output\citep{2015PhR...561....1K}. These events are hypothesized to arise from the cataclysmic collapse of massive stars, resulting in the formation of a black hole \citep{2006ARA&A..44..507W}. Conversely, short GRBs, which account for about 30\% of all GRBs, exhibit durations of less than 2 seconds and lower energy output compared to long GRBs \citep{2005Natur.437..851G}. The prevailing model for the origin of short GRBs involves the merger of two neutron stars \citep{1989Natur.340..126E}, which has been substantiated by the occurrence of the gravitational event GW\,170817 \citep{2017PhRvL.119p1101A} accompanied by electromagnetic counterpart GRB\,170817A/AT\,2017gfo \citep[e.g.,][]{2017Natur.551...64A, 2017Sci...358.1556C, 2017ApJ...848L..17C, 2017Sci...358.1559K, 2017ApJ...848L..16S}.

Since the discovery of GRBs over 50 years ago, the understanding of their radiation mechanism and emission region has remained incomplete. The fireball model, a prevailing theory, posits that the prompt emission consists of two components: a thermal component emitted when the jet becomes optically thin (known as the photosphere region), and a non-thermal component generated by collisions between shells of varying velocities within the jet, in a region distant from the progenitor star. The radiation mechanism responsible for this emission is synchrotron radiation and synchrotron self-Compton scattering of relativistic electrons. The fireball model predicts that the prompt emission should be dominated by a thermal component \citep{1986ApJ...308L..47G}. However, observations have shown that the majority of gamma-ray radiation from gamma-ray bursts is non-thermal, often described by a smoothly joined broken power-law known as the Band function \citep{1993ApJ...413..281B}. With the advent of the \textit{Fermi} era, an increasing number of studies have revealed the presence of additional components alongside the non-thermal Band components in some GRBs, such as GRB\,100507 \citep{2013MNRAS.432.3237G}, GRB\,100724B \citep{2011ApJ...727L..33G}, GRB\,101219B \citep{2015ApJ...800L..34L}, GRB\,110721A \citep{2012ApJ...757L..31A} and GRB\,120323A \citep{2013ApJ...770...32G}. Some bursts even exhibit a dominance of thermal components, as seen in the cases of GRB\,090902B and GRB\,220426A \citep{2022ApJ...940..142W}. \cite{2023arXiv230311083W} discovered a decaying thermal component in the second brightest gamma-ray burst event, GRB\,230307A, which also displayed a broken ``$\alpha$-intensity" behavior in the energy spectrum. \cite{2021ApJS..254...35L} conducted a comprehensive analysis of a large sample of multi-pulse gamma-ray bursts observed by \textit{Fermi}, revealing that most gamma-ray bursts exhibited photosphere components in the early stages of radiation, close to the trigger, while the late period was dominated by non-thermal components.

In this paper, we present the analysis of the GRB\,200613A, and find that there is a thermal component in the prompt phase. Additionally, we conduct a detailed investigation of the multi-band afterglow and the underlying massive host galaxy. The structure of this paper is as follows: Section 2 provides a description of the observation and data reduction procedures. In Section 3, we introduce our modeling approach for both the prompt emission and afterglow emission. The discussion and conclusion are presented in Section 4 and Section 5, respectively. Throughout this study, we adopt a conventional cosmological model with the following parameter values: $H_0 = 69.6$ km s$^{-1}$ Mpc$^{-1}$, $\Omega_M = 0.286$, and $\Omega_{\Lambda} = 0.714$ \citep{2014ApJ...794..135B}.

\section{Observations and Data Redoction}

\subsection{Space-base Observations}

GRB\,200613A was triggered by Gamma-ray Burst Monitor (GBM) onboard the Fermi Gamma-ray Space Telescope (\textit{Fermi} hereafter) on June 13, 2020 at 05:30:08.214UT ($T_0$), with duration $T_{90}\sim 470$\,s (50-300keV) and fluence $(4.10\pm 0.05) \times 10^{-5}\mbox{erg/cm}^2$ in 10-1000keV \citep{2020GCN.27930....1B}. The prompt emission exhibited a double peak profile with a long, weak emission tail that persisted until $\sim 500$s. The Large Area Telescope (LAT) of Fermi also detected the burst after the GBM trigger, with a photon flux and photon index higher than 100MeV of $(6.6\pm3.6)\times 10^{-7}$ph/cm$^2$/s and $-1.73 \pm 0.36$, respectively \citep{2020GCN.27931....1O}. 

Subsequently, the X-ray Telescope (XRT) onboard \textit{Neil Gehrels Swift} Observatory \citep[\textit{Swift} hereafter;][]{2004ApJ...611.1005G} began to follow up on this burst at $T_0+43.6$ ks after the Fermi/GBM trigger and identified seven uncatalogued X-ray sources within the Fermi/LAT error circle. Among these sources, ``Source 2" was found to be above the ROSAT ALL-SKY SURVEY (RASS) limit and is, therefore, the most likely GRB afterglow. The position of ``Source 2" was reported as RA = 10:12:10.08, Dec = +45:45:14.7 (J2000) with an error radius of 1.4 arcsec (90\% confidence). Preliminary analysis indicates that the spectral index of 1.89 was obtained by fitting XRT spectra with an absorbed power-law. The best fit for the absorption column was $(2.3 \pm 0.5) \times 10^{21}$ cm$^{-2}$, which exceeds the Galactic value of $8.4 \times 10^{19}$ cm$^{-2}$ \citep{kennea2020grb}.

The Ultraviolet/Optical Telescope (UVOT) slewed to the position of the burst at $T_0+43.6$ks after the Fermi/GBM trigger and detected a fading source within the XRT error region \citep{marshall2020grb}. A series of images were obtained in the U, V, UVW2, and White filters, but the optical afterglow was only detected in the White and U bands.

\subsection{Fermi Data Reduction}

The Fermi/GBM payload carries 12 sodium iodide (NaI, 8keV-1MeV) and 2 bismuth germanate (BGO, 200keV-40MeV) scintillation detectors \citep{2009ApJ...702..791M}. Considering the pointing direction of detectors, we employed 4 NaI detectors (n0, n1, n3, n9) and 1 BGO detector (b0) to conduct the spectral analysis. We obtained the Time-Tagged Event (TTE) data covering the time range of this GRB from the Fermi/GBM public data archive \footnote{\url{https://heasarc.gsfc.nasa.gov/FTP/fermi/data/gbm/daily/}}. We employed the Multi-Mission Maximum Likelihood Framework (\texttt{3ML},\cite{3ml}) as the main tool to do analyses of Fermi/GBM data. 

The Fermi/LAT extended type data for GRB\,200613A is downloaded from the Fermi Science Support Center \footnote{\url{https://fermi.gsfc.nasa.gov}}. We analyzed the Fermi/LAT data within a $12^{\circ} \times 12^{\circ}$ region of interest (ROI) centered on the position of GRB\,200613A use Conda fermitools v2.2.0 package, distributed by the Fermi Collaboration\footnote{\url{https://github.com/fermi-lat/Fermitools-conda/wiki}}. Data were extracted with the energy range from 100 MeV to 30 GeV, selected by the `TRANSIENT' class (evclass=16, $P8R3\_TRANSIENT020\_V3$), as well as limited with a maximum zenith angle of 90 degrees to reduce the contamination from the $\gamma$ ray Earth limb. The time range of the data is from 0 to 700\,s after the GBM trigger time $T_0$.  We used the corresponding isotropic emission template, \rm{iso\_P8R3\_TRANSIENT020\_V3\_v1}, as well as the diffuse Galactic interstellar emission template (IEM) \rm{gll\_iem\_V07} for the Galactic diffuse emissions. The parameters of isotropic emission and IEM are left free. The light curve of prompt emission is shown in Figure \ref{Fig.total_lc}.

\subsection{Photometry Observation and Data Reduction}
\begin{figure}[ht!]
\plotone{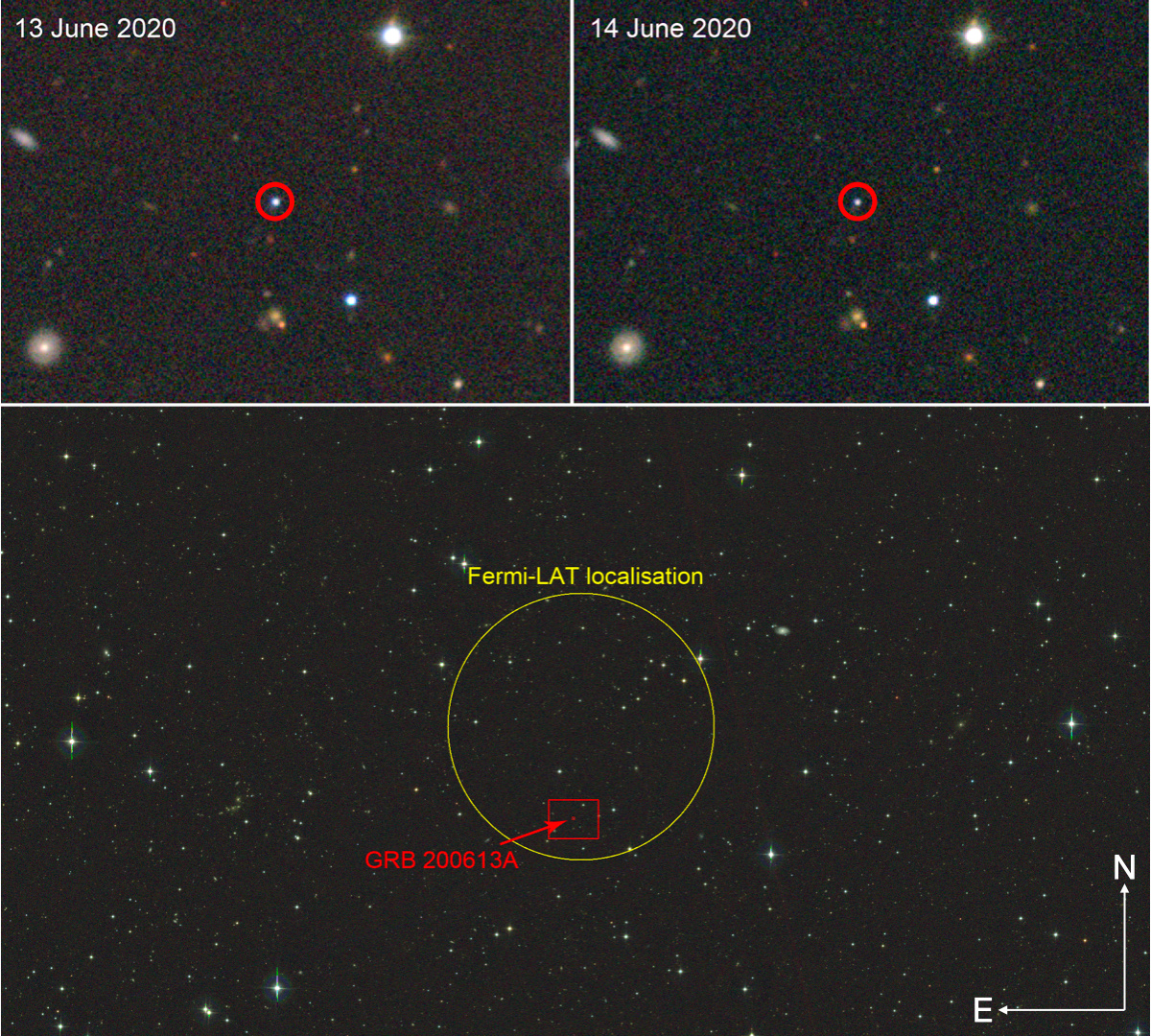}
\caption{Color composites created combining \textit{g}, \textit{r} and \textit{i} images obtained with OAJ T80. The bottom image, with a field of view of 1.2$\times$0.6 degrees, shows the location of the afterglow within the LAT error box. The top panels show the area of the images with the location of the afterglow on the first and second nights and have a field of view of 3.0$\times$2.1 arcmin. The optical afterglow is located at coordinates RA = 10:12:10.029 and Dec = +45:45:13.925 (J2000). \label{fig.oaj}}
\end{figure}

The optical afterglow of the GRB\,200613A was first detected at the position of the second XRT source within the Fermi-LAT localization, using the T80 0.8m telescope at the Observatorio de Javalambre in Spain. The optical afterglow is located at coordinates RA = 10:12:10.029 and Dec = +45:45:13.925 (J2000) with uncertainty of 1.0 arcsec. Figure \ref{fig.oaj} shows a color composite image of the optical afterglow of the GRB obtained with the T80 0.8m telescope. The image is composed of g, r, and i band frames. The Xinglong-2.16m telescope (Xinglong observatory, Hebei, China) equipped with the Beijing Faint Object Spectrograph and Camera (BFOSC) conducted follow-up observation on June 14, 2020, at 12:44:18UT, approximately 1.31 days after the GBM trigger. A series of R-band frames, each with an exposure time of 300 seconds, were acquired. The stack image clearly reveals the presence of the GRB afterglow with a magnitude of $19.76\pm 0.06$. Follow-up observation continued over the following days until the telescope reached its sensitivity limit. On June 21, 2020, we employed the 2.56 m Nordic Optical Telescope (NOT; Roque de los Muchachos observatory, La Palma, Spain) equipped with Alhambra Faint Object Spectrograph and Camera (ALFOSC) to further observe GRB\,200613A. The object remained detectable in the image, and its brightness exhibited a power-law decay with an index of approximately -0.8, as expected. Our follow-up observations with the NOT were continued until July 1. Additional observaion in SDSS-g, i, and r bands was conducted using CAFOS on the 2.2m telescope at the Calar Alto Observatory (CAHA) in Almeria, Spain, on June 15, 27, and July 2, 2020.

The ground-based optical data were processed using standard procedures with the Image Reduction and Analysis Facility (IRAF) v2.16 \citep{1986SPIE..627..733T}. This involved bias subtraction, flat-fielding, and image combination. Aperture photometry was performed on stacked frames, and the flux was calibrated using nearby Pan-STARRS1 field stars \citep{2016arXiv161205560C}. 

The Swift/UVOT sky images were obtained from the UK Swift Science Data Centre\footnote{\url{https://www.swift.ac.uk/swift\_portal/}}. The \texttt{UVOTPRODUCT} tool of Xspec was used to extract the object magnitude. The data were automatically rebinned to ensure a signal-to-noise ratio $S/N >3$. 

In addition to our own observations, we supplement our dataset with photometric data from the GRB Coordinates Network (GCN)\footnote{\url{https://gcn.nasa.gov/circulars}}. These contributions include observations from the Kitab-ISON RC-36 and AS-32 telescopes at the Abastumani observatory \citep{pozanenko2020grb, 2020GCN.27958....1B}, the Liverpool Telescope located in La Palma \citep{izzo2020grb}, the MITSuME 50 cm telescope at the Akeno Observatory \citep{murata2020grb}, the AZT-33IK 1.5-m telescope at the Sayan observatory \citep{2020GCN.27978....1B}, the Zeiss-1000 telescope at the SAO RAS \citep{moskvitin2020grb}, the 2-m Himalayan Chandra Telescope located at IAO, Hanle, India \citep{dutta2020grb}, and the ZTSH 2.6m telescope at the CrAO observatory \citep{2020GCN.28000....1B, 2020GCN.28003....1B}. All the optical photometry data are presented in Table \ref{tb.phot}. It is important to note that the magnitudes were not corrected for the Galactic foreground extinction of $E(B-V) = 0.008$ in the direction of the burst, which corresponds to $A_{V,MW}=0.024$ \citep{schlafly2011measuring}.

\subsection{Spectroscopy Observation of the Host Galaxy}

\begin{figure}[ht!]
\plotone{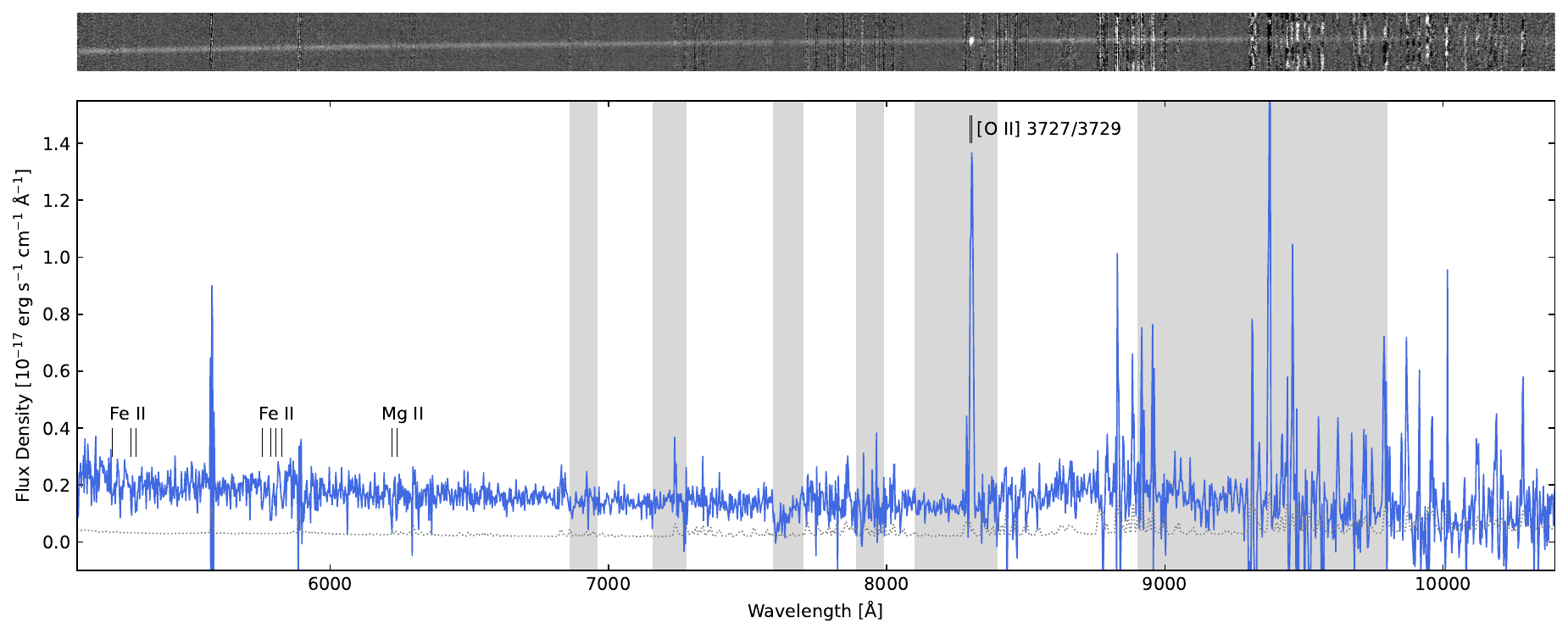}
\caption{One- and two-dimensional spectrum of the host galaxy of GRB,200613A. The blue solid line shows the spectrum, while the gray dotted line shows the error spectrum. The black vertical lines mark the positions of both emission and absorption lines. The gray region marks the region of terrestrial absorption lines. According to prominent emission lines [OII] 3727/3729 doublet, we get a redshift of $z = 1.2277 \pm 0.0011$ for GRB\,200613A, and a similar redshift of $z = 1.2262 \pm 0.0009$ for the intervening system by the absorption line. \label{Fig.host_spec}}
\end{figure}

\begin{figure}[ht!]
\plotone{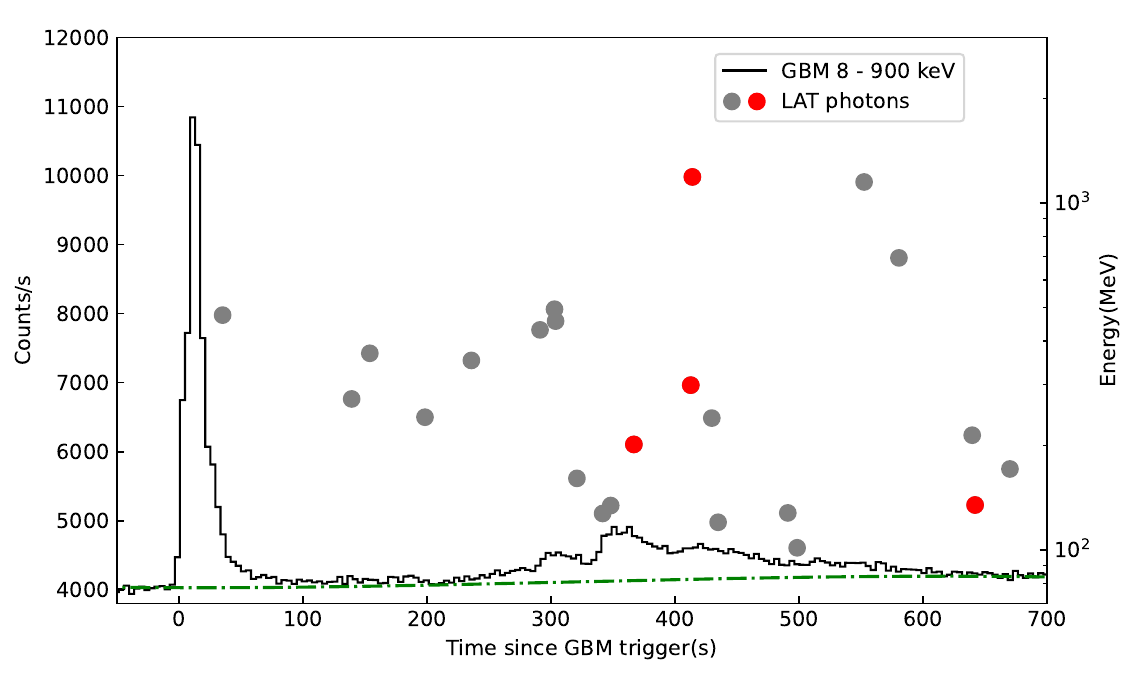}
\caption{Fermi/GBM and Fermi/LAT observational data for GRB\,200613A. The points show the arriving times and energies of Fermi/LAT photons $\ge$ 100MeV, and the photons whose probabilities associated with GRB\,200613A excess 90\% are highlighted by red. A third-order polynomial fit to the background level is shown as a green dot-dashed line.} \label{Fig.total_lc}
\end{figure}

To get the spectrum of the host galaxy of GRB\,200613A, we perform spectroscopy observation with OSIRIS, mounted on the 10.4m GTC telescope (Roque de los Muchachos Observatory, La Palma, Spain) at 23:28:10 UT on 18 January 2021 (218.73 days after the burst). We got $3\times 1200$\,s frame with grism R1000R, covering wavelength range from 5100 to 10100 \AA. The spectrum is dominated by continuum. A strong emission feature is found in $\sim8305$\,\AA which we identify as the [OII] 3727/3729 doublet with the redshift 1.228. At the blue part of the spectrum, we also find some weak absorbtion feature of FeII and MgII 2796/2803 at similar redshift 1.226. The reduced One- and two-dimensional spectrum are shown in Figure \ref{Fig.host_spec}.

The spectral analysis of the host galaxy associated with GRB\,200613A employed a one-dimensional Gaussian profile fitting procedure to model both emission and absorption lines. This approach facilitated the precise measurement of the redshift for both components. For the prominent [OII] doublet at 3727/3729 Å, a redshift of $z = 1.2277 \pm 0.0011$ was determined. Furthermore, analysis of the absorption lines revealed a redshift of $z = 1.2262 \pm 0.0009$. Taking these results into consideration, we conclude that the host galaxy of GRB 200613A resides at a redshift of $z = 1.2277 \pm 0.0011$. The observed absorption lines are likely attributed to an intervening system located at a similar redshift of $z = 1.2262 \pm 0.0009$, suggesting the presence of intervening material along the line of sight towards the GRB.

\startlongtable
\begin{deluxetable*}{ccccc}
\tablecaption{The multi-band photometry results of GRB\,200613A and publicly available data from Gamma-ray Coordination Network (GCN).}
\tablewidth{0pt}
\label{tb.phot}
\tablehead{
\colhead{$\Delta T$} & \colhead{Band} & \colhead{Magnitude} & \colhead{Telescope}& \colhead{Ref.} \\ 
\colhead{(day)} & \colhead{}  & \colhead{(AB)} & \colhead{Inst.}& \colhead{}
}
\startdata
        1.31 & R & $19.69 \pm 0.05$ & XL216/BFOSC & this work \\ 
        2.31 & R & $20.22 \pm 0.09$ & XL216/BFOSC & this work \\ 
        3.32 & R & $20.71 \pm 0.16$ & XL216/BFOSC & this work \\ 
        5.32 & R & $21.1 \pm 0.16$ & XL216/BFOSC & this work \\ 
        6.32 & R & $20.97 \pm 0.15$ & XL216/BFOSC & this work \\ 
        8.3 & R & $ > 21.30$ & XL216/BFOSC & this work \\ 
        8.66 & r & $21.53 \pm 0.08$ & NOT/ALFOSC & this work \\ 
        12.67 & r & $21.84 \pm 0.08$ & NOT/ALFOSC & this work \\ 
        18.66 & r & $ > 21.67$ & NOT/ALFOSC & this work \\ 
        18.68 & i & $21.73 \pm 0.23$ & NOT/ALFOSC & this work \\ 
        10.69 & g & $21.64 \pm 0.10$ & Liverpool Telescope & this work \\ 
        10.69 & r & $21.59 \pm 0.10$ & Liverpool Telescope & this work \\ 
        10.7 & i & $21.19 \pm 0.11$ & Liverpool Telescope & this work \\ 
        2.64 & g & $20.86 \pm 0.19$ & CAHA/CAFOS & this work \\ 
        3.68 & g & $21.37 \pm 0.11$ & CAHA/CAFOS & this work \\ 
        6.64 & g & $21.49 \pm 0.06$ & CAHA/CAFOS & this work \\ 
        2.65 & r & $20.51 \pm 0.10$ & CAHA/CAFOS & this work \\ 
        3.69 & r & $20.92 \pm 0.08$ & CAHA/CAFOS & this work \\
        6.67 & r & $21.17 \pm 0.05$ & CAHA/CAFOS & this work \\ 
        19.68 & r & $>21.54$ & CAHA/CAFOS & this work \\ 
        2.66 & i & $20.53 \pm 0.23$ & CAHA/CAFOS & this work \\ 
        3.70 & i & $20.59 \pm 0.11$ & CAHA/CAFOS & this work \\
        6.68 & i & $20.99 \pm 0.06$ & CAHA/CAFOS & this work \\ 
        0.69 & g & $19.62 \pm 0.02$ & OAJ-T80 & this work\\
        1.71 & g & $20.41 \pm 0.04$ & OAJ-T80 & this work\\
        0.68 & r & $19.35 \pm 0.02$ & OAJ-T80 & this work\\
        1.70 & r & $20.16 \pm 0.03$ & OAJ-T80 & this work\\
        0.67 & i & $19.19 \pm 0.02$ & OAJ-T80 & this work\\
        1.69 & i & $19.99 \pm 0.04$ & OAJ-T80 & this work\\
        0.55 & R & $19.27 \pm 0.07$ & \nodata & \cite{2020GCN.27937....1P} \\ 
        0.71 & r & $19.45 \pm 0.20$ & Liverpool Telescope & \cite{2020GCN.27939....1I} \\ 
        2.53 & R & $20.47 \pm 0.11$ & AS-32 Telescope & \cite{2020GCN.27958....1B} \\ 
        3.58 & R & $20.58 \pm 0.05$ & \nodata & \cite{2020GCN.27982....1M} \\ 
        4.47 & R & $20.94 \pm 0.14$ & \nodata & \cite{2020GCN.27978....1B} \\ 
        6.41 & R & $20.59 \pm 0.26$ & \nodata & \cite{2020GCN.27999....1D} \\ 
        6.58 & R & $21.06 \pm 0.09$ & \nodata & \cite{2020GCN.28000....1B} \\ 
        6.77 & r & $21.20 \pm 0.30$ & \nodata & \cite{2020GCN.27998....1I} \\ 
        7.93 & R & $21.13 \pm 0.10$ & \nodata & \cite{2020GCN.28003....1B} \\ 
        0.62 & white & $20.18 \pm 0.05$ & Swift/UVOT & this work \\ 
        0.78 & white & $20.33 \pm 0.13$ & Swift/UVOT & this work \\ 
        3.07 & white & $21.73 \pm 0.12$ & Swift/UVOT & this work \\ 
        3.53 & white & $21.58 \pm 0.28$ & Swift/UVOT & this work \\ 
        6.03 & white & $22.01 \pm 0.14$ & Swift/UVOT & this work \\ 
        8.15 & white & $21.99 \pm 0.16$ & Swift/UVOT & this work \\ 
        10.04 & white & $22.44 \pm 0.25$ & Swift/UVOT & this work \\ 
        10.34 & white & $ > 22.04$ & Swift/UVOT & this work \\ 
        14.16 & white & $22.46 \pm 0.20$ & Swift/UVOT & this work \\ 
        0.54 & U & $19.89 \pm 0.12$ & Swift/UVOT & this work \\ 
        0.68 & U & $19.87 \pm 0.09$ & Swift/UVOT & this work \\ 
        0.78 & U & $19.97 \pm 0.13$ & Swift/UVOT & this work \\ 
        0.59 & V & $ > 19.05$ & Swift/UVOT & this work \\ 
        4.03 & UVW2 & $ > 23.51$ & Swift/UVOT & this work \\ 
\enddata
\tablecomments{The photometry results presented in this table have not been corrected for Galactic extinction and have not taken into account the potential contamination of host galaxy emission. }
\end{deluxetable*}

\section{Results}
\subsection{Prompt Emission} \label{prompt_emission_results}
\subsubsection{Evolution of Spectral Index}\label{Sec.Ep}
The prompt emission of GRB\,200613A exhibits a bright main multi-structure pulse followed by a week-long emission tail with duration $T_{90} = 478.03\pm3.17$\,s (in 50-300 keV). We select a time interval from $T_0-0.32$ to $T_0+37.96$\,s to build time-average spectra for the main pulse using different models, namely the Band, Cutoff power-law, and Blackbody (BB). Our analysis reveals that the Band function can well describe the spectrum, with the best-fit results of $\alpha = -1.16_{-0.03}^{+0.03}$, $\beta = -3.20_{-0.36}^{+0.29}$, and $E_p = 116.0_{-4.4}^{+4.6}$ keV. Combined with the redshift of the burst $z=1.2277$, we calculated the isotropic energy of the main pulse of prompt emission $E_{\gamma,iso}=1.68_{-0.06}^{+0.07}\times10^{53}$ erg.

\begin{figure}[ht!]
\plotone{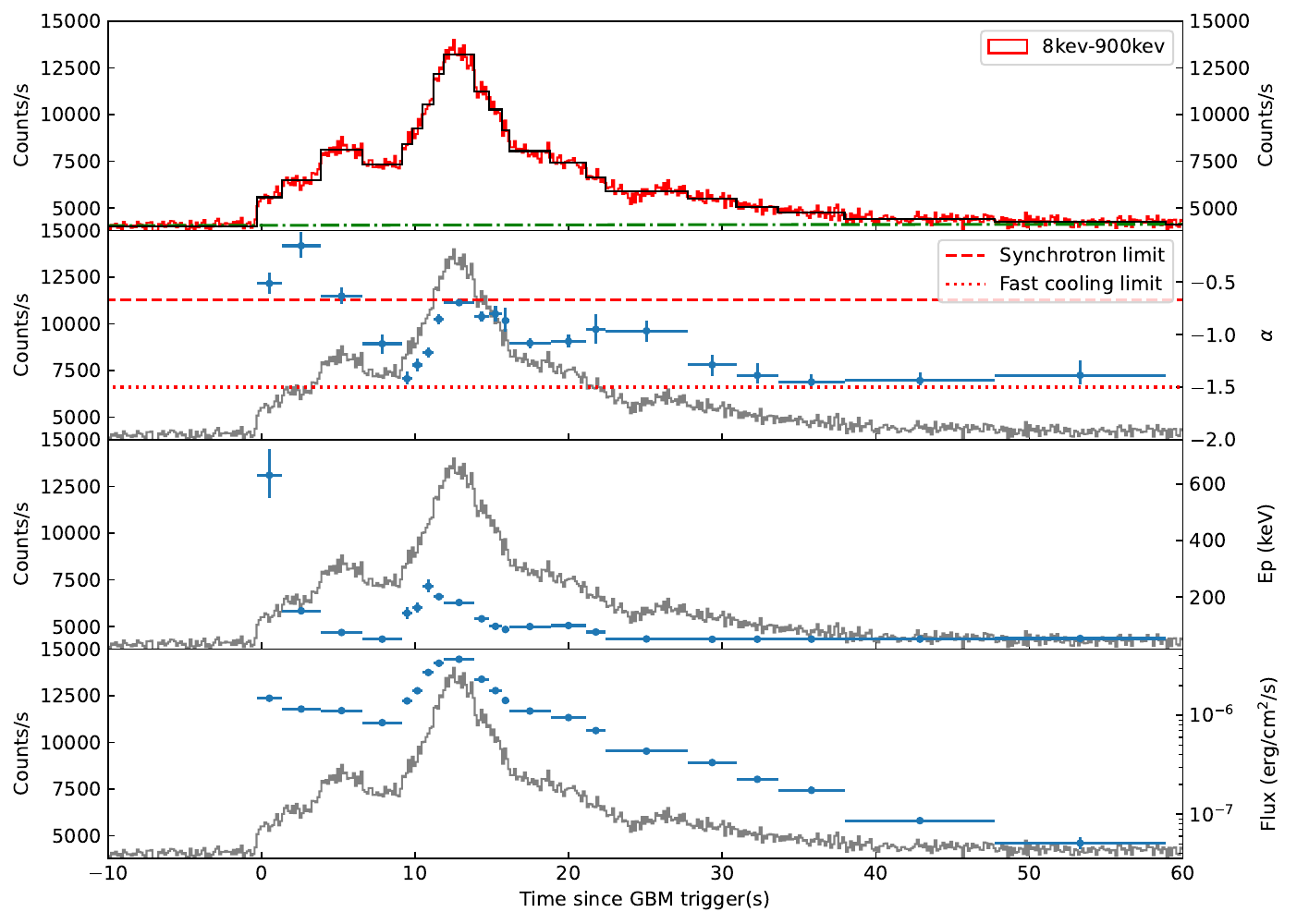}
\caption{Observational data and parameter evolution of the main pulse of GRB\,200613A. The light curve (red) of the main pulse with a time resolution of 0.128s and the Bayesian block (black) are shown in the first panel. The background level is indicated by the green dot-dashed line. The subsequent three panels display the evolution of the spectral index ($\alpha$), peak energy ($E_p$), and energy flux (10 - 1,000 keV) in the Band model. In the second panel, the synchrotron limit (-2/3) and fast cooling limit (-3/2) are illustrated by a dashed line and a dotted line, respectively. \label{Fig.prompt_sp}}
\end{figure}

To further investigate the spectral evolution of the prompt emission, we conduct time-resolved spectral analysis for the main pulse using the Bayesian block method to determine statistically optimal time bins. Each binned spectrum was then fitted using various models, including the Band, BB, powerlaw+BB, and Band+BB models. The results of these fits are presented in Table \ref{tb.prompt_sed}. In Figure \ref{Fig.prompt_sp}, we plot the evolution of the spectral index derived by Band model and overlay it with the GBM light curve. The dashed horizon line and dotted line in the second panel represent the synchrotron ``Line of Death" index (-2/3) and the synchrotron fast-cooling spectral index (-3/2), respectively. It's evident that the low-energy index $\alpha$ between $T_0$ to $T_0+6$\,s exceeds the ``Line of Death". However, after $T_0+7$ s, it drops within the allowed region of the synchrotron and exhibits a ``flux-tracking" pattern. The peak energy also shows a similar transition pattern. 

\begin{figure}[ht!]
\plotone{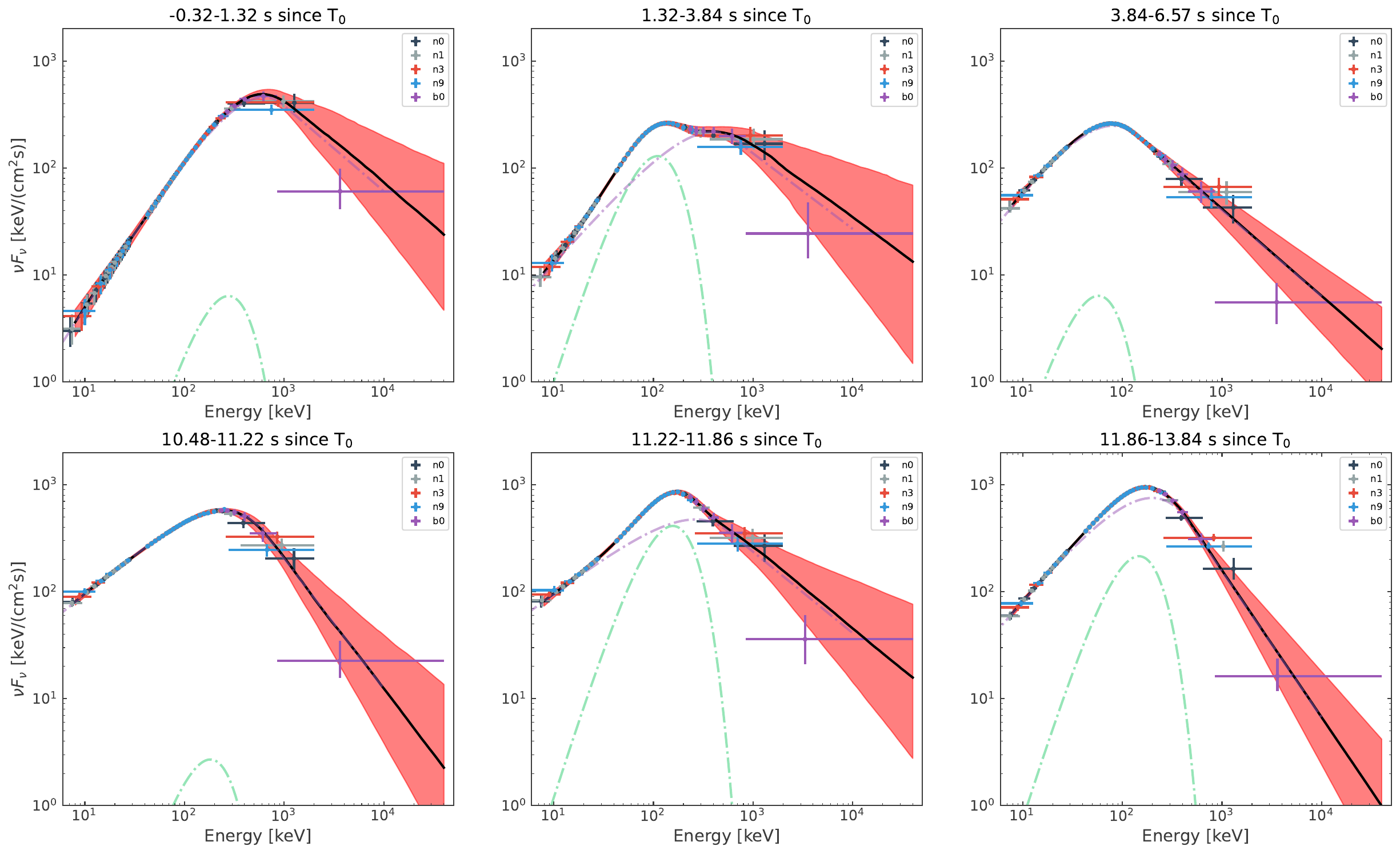}
\caption{The best fit spectral energy distribution (SED) of the Gamma-ray Burst Monitor (GBM) data using a combination of the Band function and blackbody model. The purple dot-dashed line and green dot-dashed represent the Band component and blackbody component, respectively. Solid dashed lines represent the $\nu F_{\nu}$ spectra resulting from the combination of these two models, and red shade show the 1-sigma error. \label{Fig.BandBB}}
\end{figure}

Based on the Bayesian Information Criterion \cite[BIC;][]{1978AnSta...6..461S}, the data generally favors the Band+BB model. We present the best-fit spectral energy distribution (SED) for the time intervals $T_0-0.32$ to $T_0+6.57$\,s and $T_0+10.48$ to $T_0+13.84$\,s in Figure \ref{Fig.BandBB}. In these intervals, a significant blackbody radiation component can be observed. However, blackbody components are not apparent in other time periods.

\subsubsection{Spectral Lag}\label{Sec.lag}

There is a time difference between high-energy and low-energy bands of GRB photons, which is called spectral lag. Spectral lag is always significant for those long duration GRBs \cite[e.g.,][]{Norris_2000, U12,Bernardini15}, while it is typically negligible for short duration GRBs \cite[e.g.,][]{Bernardini15, Xiao2022}.

\begin{figure}[ht!]
\plotone{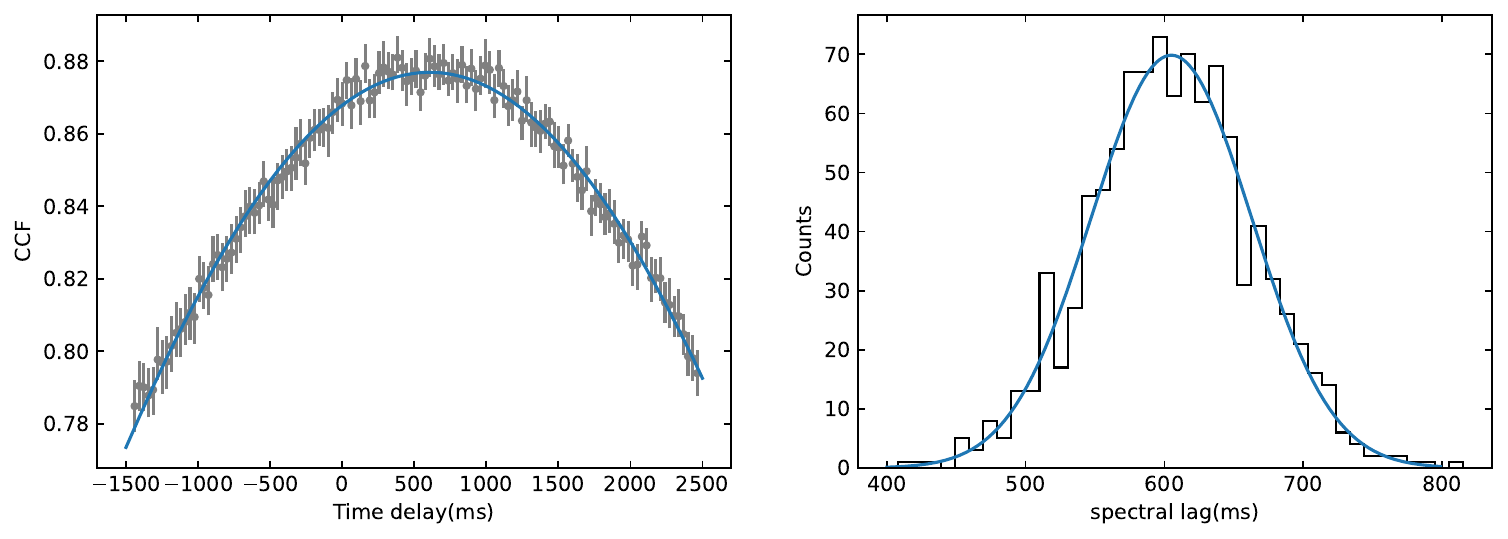}
\caption{Left panel: cross-correlation function (CCF) vs. time delay between the $(100-150)/(1+z)$ keV and $(200-250)/(1+z)$ keV channels for the main pulse. The error in each bin of the CCF is determined by running the Monte Carlo simulation 1000 times. The blue solid line shows a Gaussian fit. The time delay corresponding to the peak of the Gaussian fit determines the spectral lag. Right panel: histogram of 1000 simulated spectral lag values for the main pulse between $(100-150)/(1+z)$ keV and $(200-250)/(1+z)$ keV. The standard deviation of this simulated spectral lag distribution is considered as its associated uncertainty. The derived spectral lag of the main pulse is $607.6 \pm 57.8$ ms.\label{Fig.lag_c}}
\end{figure}

We calculated the spectral lag of GRB\,200613A by using cross-correlation function \citep[CCF;][]{Band97,Norris_2000,U10} method between (100-150)/(1+z) keV and (200-250)/(1+z) keV. For the uncertainty of lags, we used the Monte Carlo simulation \citep[see][]{Peterson98,U10}. The lag of the main pulse is $607.6 \pm 57.8$ ms. The results of the lag in 32 ms time bins of light curves and its uncertainty are illustrated in Fig. \ref{Fig.lag_c}. Over the entire duration of the burst, the spectral lag measures $614.7 \pm 56.1$ ms, which is in good agreement with the observed lag in the main pulse.

\subsubsection{Minimal Variable Time Scale}\label{MVTS}

The variable central engine emits relativistic outflow consisting of multiple shells with different velocities \citep{1994ApJ...430L..93R}. The kinetic energy converts to radiation when the faster shell hits the slower shell ahead, resulting in the production of a complex light curve in the prompt emission phase. The rapid variability of prompt emission within a short period of time provides insights into the dimensions of the emission region. The minimum variability time scales ($t_{\mbox{mv}}$) can be used to constrain the size of the minimum emission region ($R_{c}$) and the minimal bulk Lorentz factor ($\Gamma_{min}$). We employ the Bayesian Blocks algorithm and define the 1/2 shortest significant structures of blocks as the duration of minimum time interval \citep{vianello18}, and the minimum variability time scales of GRB\,200613A is $t_{\mbox{mv}} = 267$ ms. Thus, we can evaluate the minimal bulk Lorentz factor and the size of the emission region ($R_c$) from the following equation \citep{2015ApJ...811...93G}:

\begin{equation}
\Gamma_{\mbox{min}}\gtrsim 110\left(\frac{L_{\gamma,iso}}{10^{51}\mbox{ erg s}^{-1}}\frac{1+z}{t_{\mbox{mv}}/0.1\mbox{ s}}\right)^{1/5}
\end{equation}

\begin{equation}
R_c \simeq 7.3\times 10^{13}\left(\frac{L_{\gamma,iso}}{10^{51}\mbox{ erg s}^{-1}}\right)^{2/5}\left(\frac{t_{\mbox{mv}}/0.1\mbox{ s}}{1+z}\right)^{3/5} \mbox{cm.}
\end{equation}
Here, for simplicity, the luminosity in the equation is estimated as $L_{\gamma,iso} = E_{\gamma,iso}/T_{90}$. The derived values for $\Gamma_{\mbox{min}}$ and $R_c$ are found to be $\Gamma_{\mbox{min}} \gtrsim 89$ and $R_c \simeq 5.7\times 10^{13}$ cm, respectively.

\subsection{Afterglow Analysis}
\subsubsection{Temporal Analysis}
\begin{figure}[ht!]
\plotone{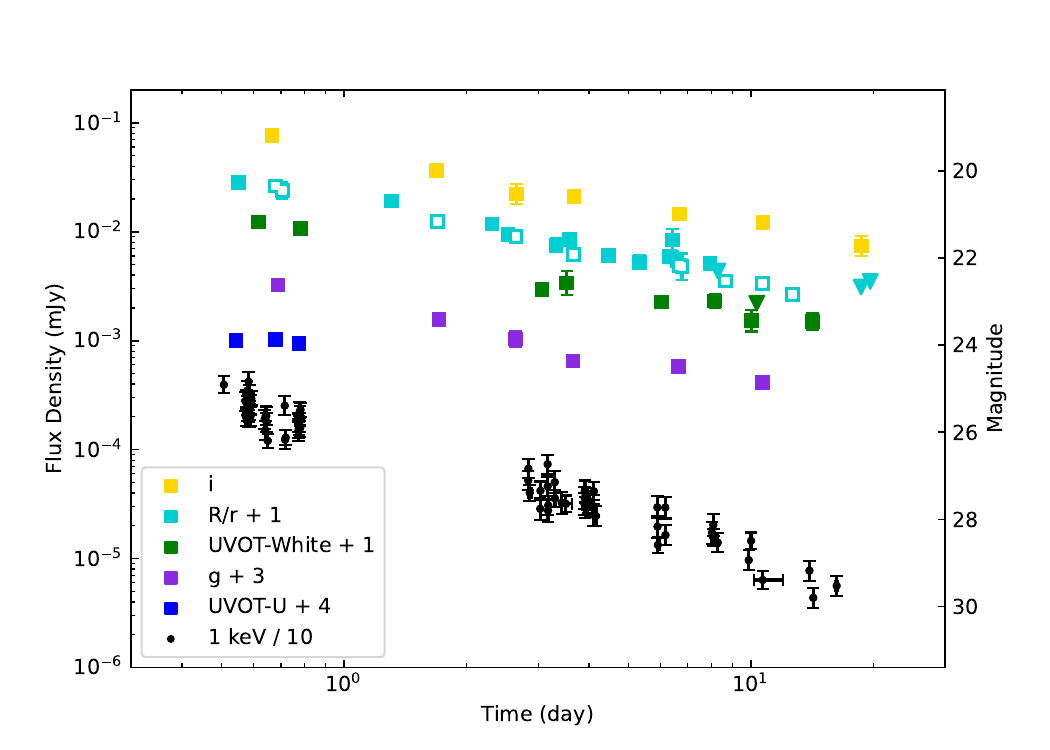}
\caption{Multi-band afterglow light curve of GRB\,200613A spanning a temporal range of 0.5 to 16.25 days post-burst. Solid (hollow) green data points represent R(r) band observations. The inverted triangle denotes the limiting magnitude. For convenience, we added offsets in fluxes as shown in the legend. Note that the data are not corrected for the Galactic extinctions and the host galaxy contribution. \label{Fig.lc}}
\end{figure}

The multi-band afterglow light curves of GRB\,200613A were obtained from $\sim$ 0.5 days to 19 days after the GBM trigger (Figure \ref{Fig.lc}). Empirical functions were employed to model the temporal profile in the optical/X-ray light curve. Given the smooth decay in the optical bands, a single-power law (SPL) was used to fit the optical data, which is expressed as
\begin{equation}
	F=F_0t^{-\alpha},
\end{equation}
where $F_0$ is the flux and $\alpha$ is the decay index. 

The X-ray light curve exhibits a discernible break around 7 days after the burst. So we use both single-power law and broken power law (BPL) models to fit the data. The BPL function is expressed
\begin{equation}
F = \begin{cases}
F_0\left(t/t_{b}\right)^{-\alpha_1},\quad &t\leq t_b \\
F_0\left(t/t_{b}\right)^{-\alpha_2},\quad &t>t_b
\end{cases} 
\end{equation}
where $t_{b}$ is the break time of light curve. The temporal fitting results are presented in Table \ref{tb.temp_fit}.

\begin{deluxetable*}{cccc}
\tablecaption{Measurements of temporal index}
\tablewidth{0pt}
\label{tb.temp_fit}
\tablehead{
\colhead{Band} & \colhead{Model} & \colhead{Parameters} & \colhead{$\chi^2 / d.o.f$}
}
\startdata
X-ray & SPL & $\alpha=1.09^{+0.02}_{-0.02}$ & 1.42\\
\cline{2-4}
 & BPL & $\alpha_1=1.04^{+0.03}_{-0.04}$ & \\
 & {} & $\alpha_2=1.51^{+0.49}_{-0.26}$ & 1.36\\
 & {} & $\log t_b=5.80^{+0.10}_{-0.31}$ & \\
\hline
White & SPL & $\alpha=0.74^{+0.04}_{-0.04}$ & 1.34 \\
\hline
R/r & SPL & $\alpha=0.76^{+0.02}_{-0.02}$ & 2.05\\
\hline
\enddata
\tablecomments{SPL: single power-law; BPL: broken power-law. The break time $t_b$ is in a unit of seconds. The jet break effect at late time in optical may be counteracted by the emission of the host galaxy.}
\end{deluxetable*}

\subsubsection{Forward Shock Modeling}
A relativistic jet moving forward collides with the circumstellar medium, resulting in a pair of forward and reverse shocks. Electrons are accelerated in the shocks, producing multi-wavelength synchrotron radiation, known as the afterglow of GRB. In the standard external shock fireball model, the behavior of an afterglow can be described by several parameters: the isotropic energy $E_{k}$, the initial Lorentz factor $\Gamma_0$, the electron spectral index $p$, the circumburst medium density $n_{18}$ (the density at $10^{18}$ cm, defined in \texttt{PyFRS}\footnote{\url{https://github.com/leiwh/PyFRS}}), jet half-opening angle $\theta_{j}$, viewing angle of observer $\theta_{obs}$, the fractions of shock energy that go to electrons $\epsilon_e$ and magnetic ﬁeld $\epsilon_B$.

We fit the multi-wavelength data of GRB\,200613A using the standard afterglow model code \texttt{PyFRS}, as described in \cite{2014ApJ...788...32W}, \cite{2016ApJ...816...20L}, \cite{2023ApJ...948...30Z} and \cite{Zhou2024}, assuming a constant interstellar medium (see Discussion in \ref{dis.CR}). Due to limitations in the available data, we were unable to constrain the observational angle effectively, thus setting $\theta_{obs}=0$. 
For the initial Lorentz factor $\Gamma_0$, we obtained its lower limit of $\sim 89$ in Section \ref{MVTS}. In the study of $\Gamma_0$ by \cite{2018A&A...609A.112G}, there were individual events with $\Gamma_0$ greater than 1000, although the results given by different models will have some differences. Thus, the prior range of $\Gamma_0$ was set to [90, 1500]. According to \cite{2022MNRAS.511.2848A}, the value of the circumstellar medium density in the ISM case has a large numerical span, from $\sim 10^{-3}$ to $\sim 10^{3}$\, cm$^{-3}$. Therefore, we set its range slightly wider to [$10^{-4}$, $10^{4}$].

\begin{figure}[ht!]
\plotone{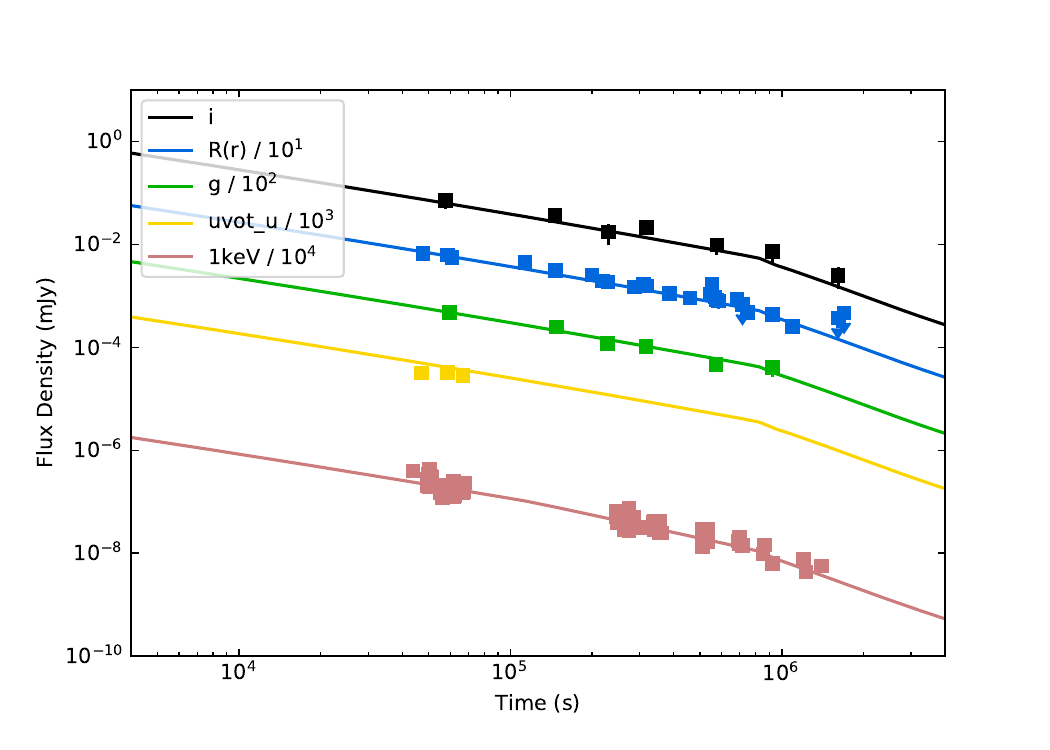}
\caption{Best fit \texttt{PyFRS} afterglow model light curve of GRB\,200613A (solid lines). All optical data plot in this figure are subtracted by host galaxy flux, with u- and i-band estimates from SDSS and g- and r-band estimates from LS (Table \ref{tb.host}). For convenience, we added offsets in fluxes as shown in the legend.
\label{Fig.LCfit}}
\end{figure}

\begin{figure}[ht!]
\plotone{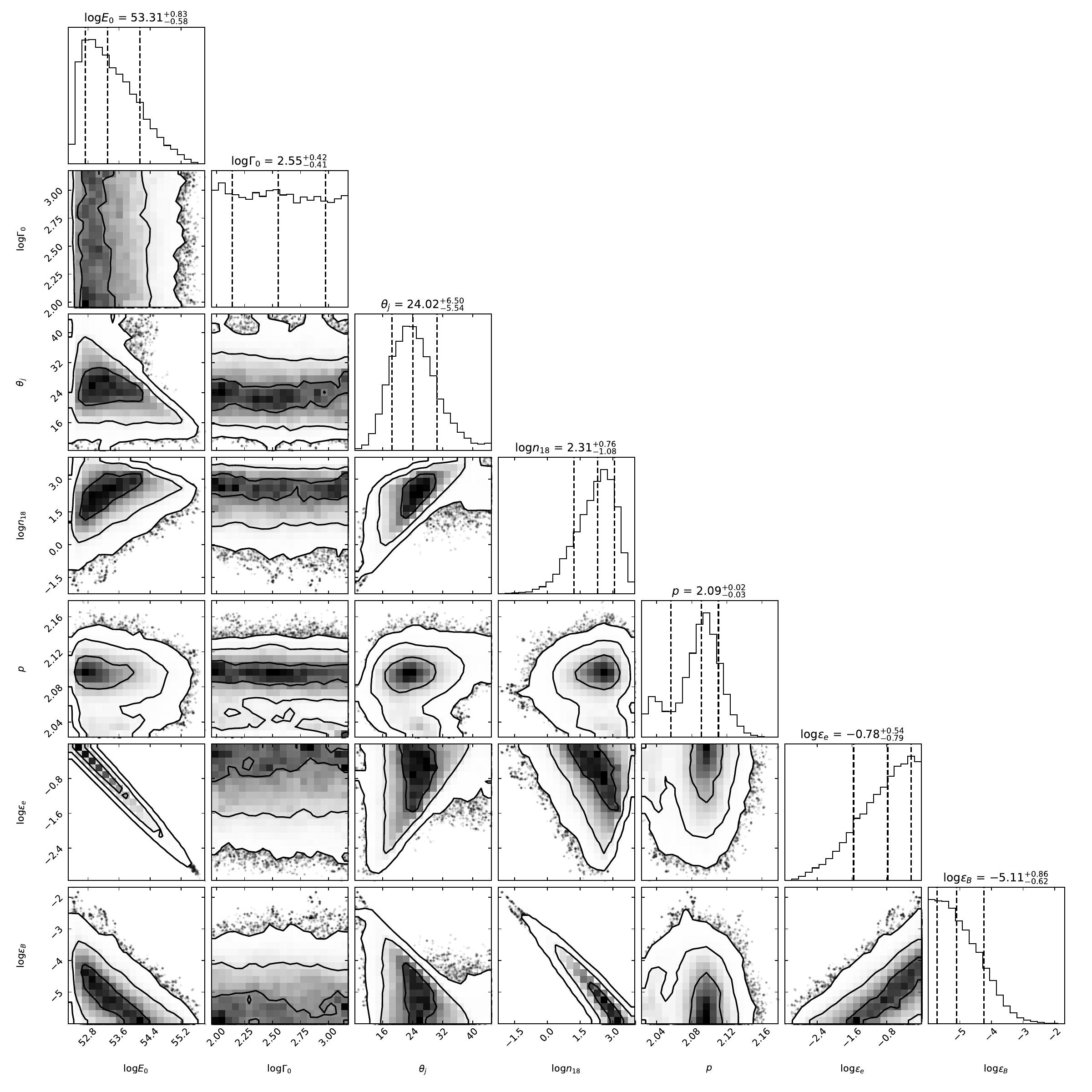}
\caption{Posterior probability distributions of the afterglow parameters of GRB\,200613A. \label{Fig.LCcorner}}
\end{figure}

The \texttt{emcee} Python package is used for performing Markov Chain Monte Carlo (MCMC) sampling to get the posterior distribution of afterglow parameters. We assume log-uniform prior distribution for $E_{\rm k}$, $\Gamma_0$, $n_{18}$, $\epsilon_{\rm e}$ and $\epsilon_{\rm B}$, and uniform prior distribution for $p$ and $\theta_{\rm j}$. We use 30 walkers to explore the parameter space with 40000 steps and discard the first 50\% of steps as burn-in. The prior range and best-fit results are listed in Table \ref{tb.AGmodel}, the best-fit light curves and posterior probability distributions of the parameters are shown in Figure \ref{Fig.LCfit} and Figure \ref{Fig.LCcorner}, respectively. 

The fitting results show that the initial Lorentz factor $\Gamma_0$ is not well constrained. This is because in order to constrain $\Gamma_0$ well, it is necessary to observe the peak time of the early optical/X-ray light curve, i.e., the deceleration timescale  
\begin{equation}
t_{\rm dec}= \left( \frac{3E_{\rm K,iso}}{16\pi n m_{\rm p} \Gamma_0^8 c^5}  \right)^{1/3},
\end{equation}
where $n$ is circumburst density, $m_p$ is the mass of proton and $c$ is speed of light. It's typically timescale is few hundred seconds. Unfortunately, for this event, the data is not sufficient to constrain $\Gamma_0$ due to the lack of early observation of the afterglow. 

At around $10^5$ seconds, a break is observed in the X-ray model light curve. Prior to this break, the X-ray and optical decay indices are consistent, while the X-ray decay index becomes slightly larger after the break. This chromatic break can be interpreted as the cooling frequency $\nu_c$ crossing the X-ray band at this time. Additionally, there is an achromatic break at around $10^6$\,s in the i, r, and X-ray bands, which is caused by the jet-break. At this time, the reciprocal of Lorentz factor is comparable to the jet angle, i.e. $1/\Gamma \sim \theta_j$. Thus, the jet half-opening angle could be well constrained to $24.0^{+6.50}_{-5.54}$ degrees.

For the circumstellar medium density $n_{18}$, although the theoretical expectation is that it may be relatively large for LGRB, the \cite{2022MNRAS.511.2848A} study shows that its distribution range is very wide for ISM. Our results are consistent with their statistical results. \cite{2014ApJ...785...29S}'s study of $\epsilon_e$ and $\epsilon_B$ found that the distribution of $\epsilon_e$ is consistent with theoretical expectations, ranging from 0.02 to 0.6, with a mean of 0.24; while the distribution of $\epsilon_B$ has a wide range, spanning 5 orders of magnitude, from $3.5\times10^{-5}$ to 0.33. Our results are also basically consistent with their statistical results.

For the uvot-u bands, the data is lower than the theoretical expectation, which we suspect is due to additional host galaxy extinction. According to the \texttt{Prospector} fitting results (see Section \ref{subsec.host}), the V-band extinction of the host is $A_V=0.28$, which means the original flux is 1.3 times the observed flux. This correction may be larger for the blue-end. Therefore, after adding additional corrections, the u-band data can roughly match the theoretical expectations.

\begin{deluxetable*}{cccc}
\tablecaption{Parameters of afterglow modeling}
\label{tb.AGmodel}
\tablewidth{0pt}
\tablehead{
\colhead{Parameters} & \colhead{Prior Type} & \colhead{Prior} & \colhead{Results} 
}
\startdata
$E_{K,iso}$  & log-uniform & $[10^{51}, 10^{56}]$ &  $(2.04^{+11.8}_{-1.50})\times 10^{53}$ erg\\
$\Gamma_0$  & log-uniform & $[90, 1500]$ &  $354^{+578}_{-217}$\\
$p$ & uniform &  $[2.01, 3.0]$ &  $2.09^{+0.02}_{-0.03}$\\
$n_{18}$ & log-uniform& $[10^{-4}, 10^4]$ & $(2.04^{+9.71}_{-1.87})\times 10^{2}$ cm$^{-3}$\\
$\theta_j$ & uniform & $[0.01, 45]$ & $24.0^{+6.50}_{-5.54}$ degrees\\
$\theta_{obs}$ & fixed & $0$ & $0$ degrees\\
$\epsilon_e$ & log-uniform & $[10^{-4}, 1]$ & $(1.66^{+4.09}_{-1.39})\times 10^{-1}$\\
$\epsilon_B$ & log-uniform & $[10^{-6}, 1]$ & $(7.76^{+48.5}_{-5.90})\times 10^{-6}$\\
\enddata
\end{deluxetable*}

\subsection{Host Galaxy}\label{subsec.host}
Large sky surveys offer a unique opportunity to study the underlying host galaxies of GRBs. We searched the Sloan Digital Sky Survey (SDSS) \citep{2023ApJS..267...44A}, Legacy Survey \citep{2019AJ....157..168D}, and The Wide-field Infrared Survey Explorer (WISE) \citep{2010AJ....140.1868W} sky survey databases and found a galaxy overlapping with the burst position of GRB\,200613A, which was confirmed to be the host galaxy of GRB\,200613A. The magnitude of each band of the host galaxy is listed in Table \ref{tb.host}.

\begin{deluxetable*}{ccc}
\tablecaption{Host galaxy magnitude in sky surveys}
\label{tb.host}
\tablewidth{0pt}
\tablehead{
\colhead{Band} & \colhead{Magnitude (AB)} & \colhead{References}
}
\startdata
u & $21.57 \pm 0.23$ & SDSS \\
g & $21.70 \pm 0.11$ & SDSS \\
r & $22.37 \pm 0.33$ & SDSS \\
i & $\sim 22.2$ & SDSS \\
z & $\sim 21.1 $ & SDSS \\
g & $22.26 \pm 0.06$ & Legacy Survey \\
r & $22.08 \pm 0.09$ & Legacy Survey \\
z & $21.43 \pm 0.09$ & Legacy Survey \\
W1 & $20.60 \pm 0.10$ & WISE \\
W2 & $21.16 \pm 0.31$ & WISE \\
W3 & $>16.5$ & WISE \\
W4 & $>14.6$ & WISE \\
\enddata
\tablecomments{SDSS: Sloan Digital Sky Survey; Legacy Survey: DESI Legacy Imaging Surveys; WISE: The Wide-field Infrared Survey Explorer.}
\end{deluxetable*}

To investigate the stellar population properties of the host galaxy, we use \texttt{Prospector} \citep{2021ApJS..254...22J} to fit the photometry results of the host. \texttt{Prospector} is a software for SED fitting to constrain the host galaxies' properties. We fit the photometric data of the host galaxy to determine key stellar population parameters, including the formed total mass ($M_F$, which refers to the cumulative mass of stars that have been formed in a galaxy over its history), the age of the galaxy at the time of observation ($t_{age}$), the stellar metallicities ($Z$), and the dust attenuation ($A_V$). During the fitting process, these parameters are allowed to vary within a prior distribution across the parameter space. To model the host galaxy, we adopt a Chabrier initial mass function (IMF) \citep{2003PASP..115..763C}, the Milky Way extinction law \citep{1989ApJ...345..245C}, and include nebular emission to account for star-forming activity. We also employ a parametric delayed-$\tau$ star formation history (SFH), which is given by the equation:
\begin{equation}
    \mbox{SFR}(t) = M_F\times \left[\int_0^t te^{-t/\tau} \mbox{d}t \right]^{-1}\times te^{-t/\tau}.
\end{equation}
Here, $\tau$ represents the e-folding time of the delayed-$\tau$ SFH. To determine the mass-weighted age of the galaxy ($t_m$), we use the observed age ($t_{age}$) and $\tau$ to calculate mass-weighted age $t_m$ using the formula \citep{2020ApJ...904...52N}:
\begin{equation}
    t_{m}=t_{age}-\frac{\int_{0}^{t_{age}} t\times \mbox{SFR}(t) \mbox{d}t}{\int_{0}^{t_{age}} \mbox{SFR}(t) \mbox{d}t}.
\end{equation}
In practical observation, the stellar mass $M_*$ is easier to obtain because it measures the mass retained by the stellar population, which is estimated by $t_m$ and $M_F$ with the following approximation \citep{2020ApJ...904...52N}:
\begin{equation}
    M_\ast \approx M_F \times 10^{1.06-0.24\log(t_m)+0.01\log(t_{m})^2}.
\end{equation}
And we derive the total dust attenuation $A_V$ from the optical depth of young stellar light $\tau_{V,1}$ \cite[assumed to be 0 here, see][]{2022ApJ...940...57N} and optical depth of old stellar light $\tau_{V,2}$:
\begin{equation}
    A_V=1.086\times (\tau_{V,1}+\tau_{V,2}).
\end{equation}

\begin{figure}[ht!]
\plotone{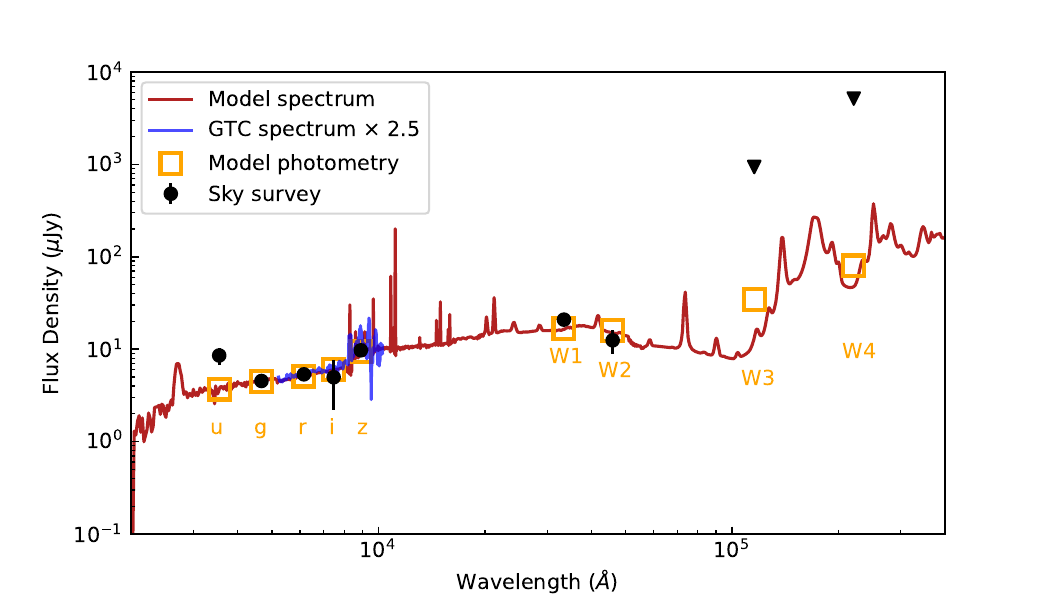}
\caption{The figure shows the SED of the host galaxy of GRB\,200613A. The best-fit model SED is shown in red line, and the slit-corrected observed GTC spectrum is shown in blue line. The model photometry results are marked as squares, and the sky survey results are represented by black data points. The g, r, and z bands magnitudes were adopted from the LS rather than SDSS due to higher S/N. All the data are corrected for the Galactic extinctions.\label{Fig.host_sed}}
\end{figure}

\begin{figure}[ht!]
\plotone{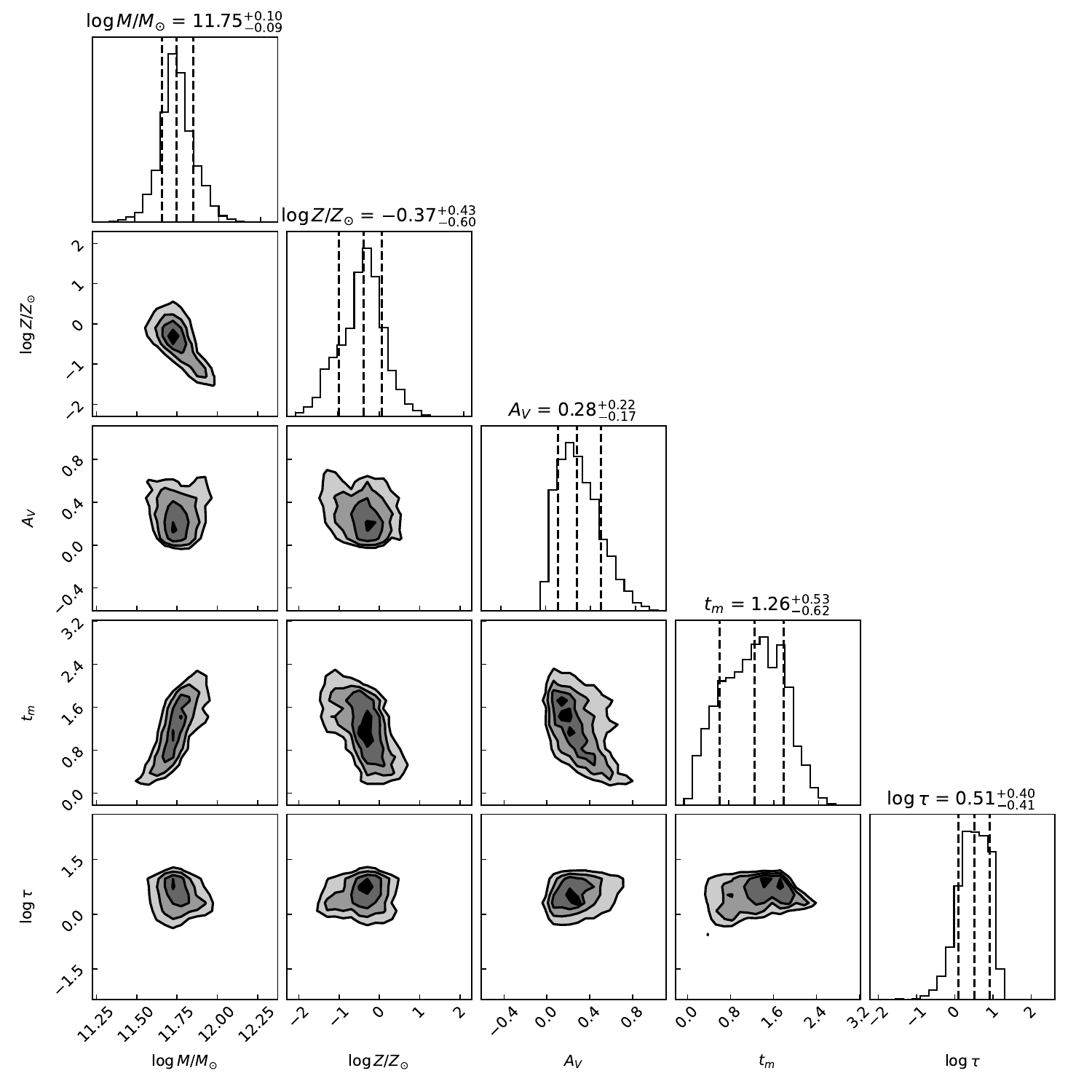}
\caption{Posterior probability distributions of the host galaxy parameters of GRB\,200613A. \label{Fig.host_corner}}
\end{figure}

During the fitting process, we adopt the the g, r, and z bands magnitudes from the LS rather than SDSS. This is because the object in the LS have a higher signal-to-noise ratio compared to those in the SDSS. Since the host galaxy is not detected in the W3 and W4 bands, we adopt the magnitudes of W3 and W4 as the typical limiting magnitudes of WISE. We derive the following parameters of host galaxy: $\log(M_\ast / M_\odot)=11.75^{+0.10}_{-0.09}$, $\log(Z / Z_\odot)=-0.37^{+0.43}_{-0.60}$, $t_{m}=1.26^{+0.53}_{-0.62}$ Gyr, $A_V=0.28^{+0.22}_{-0.17}$ mag, $\log(\tau)=0.51^{+0.40}_{-0.41}$ ($\tau$ is in unit of Gyr) and $\mbox{SFR}=22.58^{+13.63}_{-7.22} M_{\odot}$/yr. The best-fit model SED is shown in Figure \ref{Fig.host_sed}, and the posterior probability distributions of parameters are shown in Figure \ref{Fig.host_corner}. The galaxy spectrum observed with GTC is also plotted in Figure \ref{Fig.host_sed}, where 2.5 is the slit-correct factor obtained by correcting the spectral flux using the photometric data in the r, i, and z bands.

\begin{deluxetable*}{cccc}
\tablecaption{Parameters and Prior Distributions of \texttt{Prospector}}
\label{tb.host_para}
\tablewidth{0pt}
\tablehead{
\colhead{Parameters} & \colhead{Definition} & \colhead{Prior Type} & \colhead{Prior} 
}
\startdata
$M_F (M_{\odot})$ & formed total mass & log-uniform & $[10^{10}, 10^{12}]$  \\
$log(Z / Z_\odot)$ & stellar metallicity &  uniform & $[-1, 1]$  \\
$\tau_{V,2}$ & optical depth of old stellar light &  uniform & $[0, 0.5]$ \\
$\tau$ (Gyr) & e-folding time of delayed-$\tau$ SFH & log-uniform & $[10^{-3}, 13.6]$ \\
$t_{age}$ (Gyr) & age of the galaxy at the time of observation &  uniform & $[0, 5.081]$ \\
\enddata
\end{deluxetable*}

The hydrogen Balmer line $H_{\alpha}$ is a direct indicator of the star formation rate (SFR) of galaxies and is commonly used to calculate the SFR of nearby galaxies. However, for galaxies at higher redshifts ($z>0.4$), $H_{\alpha}$ falls outside the visible band. In the redshift range of $z = 0.4–1.5$, the most common emission lines are the [OII] 3737 and 3729 doublet. Although [OII] is not a direct indicator of SFR, it can also be used to indicate the SFR of galaxies after calibration \citep[e.g.][]{1989AJ.....97..700G, 1998ARA&A..36..189K, 2002MNRAS.332..283R, 2004AJ....127.2002K}. We estimated the SFR using the method described by \cite{2004AJ....127.2002K}:
\begin{equation}
    SFR(M_{\odot}yr^{-1})=(9.53\pm 0.91)\times 10^{-42}L([OII])\mbox{ erg s}^{-1}
\end{equation}
where L([OII]) is the observed, extinction-corrected luminosity. For the GTC spectra corrected for Galactic extinction, we fit its [OII] line with a Gaussian profile, and obtain the observed [OII] luminosity of $2.47\times 10^{42}$ erg/s. Following \cite{2009ApJ...691..182S}, we adopt a mean visual extinction $\langle A_V\rangle = 0.53$, and assume the MW extinction law, to obtain a SFR correction factor of 2.1, resulting in a host extinction-corrected SFR of 46.4 $M_{\odot}$yr$^{-1}$. If we use the extinction value $A_V = 0.28$ derived from \texttt{Prospector}, we obtain a correction factor of 1.5, resulting in a SFR of 34.2 $M_{\odot}$ yr$^{-1}$. Both of these results are in basically agreement with the SFRs derived from \texttt{Prospector}.

\section{Discussion}

\subsection{Classification of GRB 200613A}

The burst duration, denoted by $T_{90}$, has often been employed as a primary criterion for classifying GRBs into Short GRBs (SGRBs) and Long GRBs (LGRBs), with distinct underlying physical origins generally associated with these temporal divisions. However, a growing number of observations requires a more refined perspective on this issue. Some discoveries, such as GRB\,060614 \citep{2007ApJ...655L..25Z}, GRB\,160410A \citep{2023MNRAS.520..613A}, and GRB\,211211A \citep{2022Natur.612..223R, 2022Natur.612..232Y}, demonstrate that GRBs with $T_{90}>2$\,s can also originate from mergers, defying the traditional SGRB classification. Conversely, GRB\,200826A \citep{2022ApJ...932....1R} exemplifies a GRB with $T_{90}<2$\,s arising from the collapse of a massive star, challenging the presumed exclusivity of LGRBs with this temporal characteristic. These observations collectively highlight the limitations of relying solely on $T_{90}$ as a definitive indicator of a GRB's physical origin. To achieve a more accurate representation of the origin for the two types of GRBs, \cite{2007ApJ...655L..25Z} proposed a classification scheme for GRBs that is analogous to the supernova classification scheme. In this scheme, GRBs are divided into two types: Type I (typically short and associated with old populations) and Type II (typically long and associated with young populations). Effectively disentangling the different mechanisms driving GRB generation necessitates the development and application of more comprehensive and multifaceted methodologies.

\begin{figure}[ht!]
\plotone{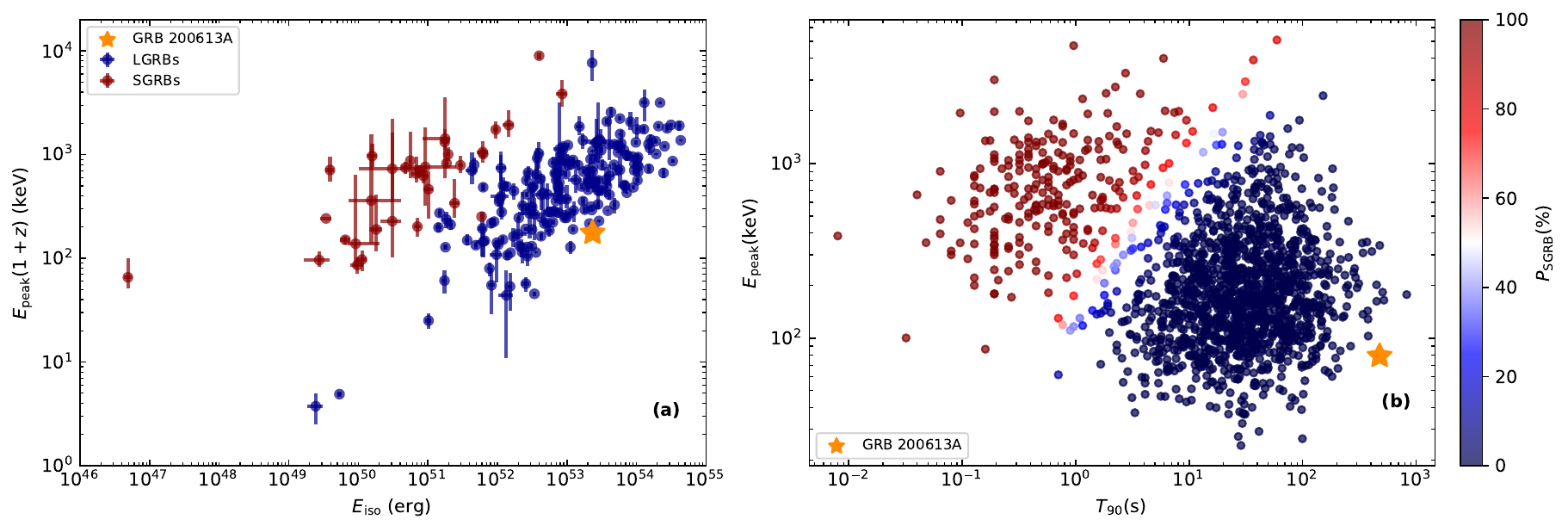}
\caption{Left panel: $E_{p,i}-E_{iso}$ diagram of GRB\,200613A (yellow star). Blue and red points represent LGRBs and SGRBs, respectively. Right panel: $E_{p}-T_{90}$ diagram of GRB\,200613A (yellow star). The color indicates the probability of being SGRB, with red being 100\% SGRB and blue being 100\% LGRB.} \label{Fig.Amati}
\end{figure}

In the field of GRB classification, the Amati relation, which links the equivalent isotropic energy $E_{\gamma,iso}$ and the intrinsic peak energy $E_{p,i}$, offers a valuable tool for differentiating between Type I and Type II bursts \citep{2002A&A...390...81A, 2006MNRAS.372..233A}. In this context, we construct time-averaged spectra for the entirety of GRB 200613A utilizing the Band model. With the measured redshift z = 1.2277, we derive the equivalent isotropic energy and peak energy to be $E_{\gamma,iso}=2.34_{-0.20}^{+0.23}\times 10^{53}$ erg and $E_p=79.0_{-11.1}^{+15.3}$ keV, respectively. Consequently, the intrinsic peak energy is $E_{p,i}=(1+z)E_{p,obs} = 176.0_{-24.7}^{+34.1}$ keV. To construct the Amati relation sample, we compiled data from three primary sources: Fermi GBM Burst Catalog \citep{2020ApJ...893...46V}, Konus-Wind Burst Catalog \citep{2017ApJ...850..161T}, and some literature \citep{2008MNRAS.391..577A, 2009ApJ...703.1696Z,  2012MNRAS.421.1256N, 2018ApJ...852L..30P, 2018ApJ...852L...1Z, 2020MNRAS.492.1919M}.  The positions of these collected samples and GRB 200613A are visualized within $E_{p,i}-E_{iso}$ plane in the left panel of Figure \ref{Fig.Amati}. Furthermore, the right panel of Figure \ref{Fig.Amati} presents the peak energy ($E_p$) and burst duration ($T_{90}$) data for 1647 Fermi GBM GRBs \citep{2020ApJ...893...46V}, where either the Comptonized function or the Band function model provided the best fit for each burst. These data are fit with two log-normal distributions, corresponding to the two distinct GRB types. The color indicates the probability of being SGRB, with red being 100\% SGRB and blue being 100\% LGRB. As evident in both plots, GRB 200613A resides on the lower-right edge of the sample distribution, indicating its classification as a typical LGRB.

Although GRB\,200613A is located on the LGRB side in the $E_p-T_{90}$ diagram, it becomes an exception due to its large $T_{90}$. This is because the prompt emission of GRB\,200613A consists of two parts: the bright main pulse radiation in the first tens of seconds, and the weak radiation that lasts for hundreds of seconds. If GRB\,200613A is located at a larger redshift, limited by the sensitivity of the detector, we can only observe the main pulse of this burst, and we will get a smaller $T_{90}$. This also means that the $T_{90}$ we observe in the GRB sample has an instrumental selection effect. For some LGRBs, the actual $T_{90}$ may be much larger than the observed value, that is, the sample will extend to the right in the $E_p-T_{90}$ diagram.

In addition, in Section \ref{Sec.lag}, we calculated the spectral lag of GRB\,200613A. Long GRBs (LGRBs) typically exhibit larger spectral lags compared to short GRBs (SGRBs), which can often be negligible. Our measured spectral lag of $614.7 \pm 56.1$ ms indecates that GRB\,200613A aligns more closely with the characteristics of an LGRB. 

The combined evidence of GRB\,200613A's long duration $T_{90}$, location on the lower-right edge of the Amati relation diagram, and significant spectral lag strongly suggests that it is a typical collapse burst (Type II burst).

\subsection{Possible Origin for the Spectral Evolution of Prompt Emission}

As shown in Section \ref{prompt_emission_results}, the spectrum of the prompt emission of GRB 200613A can be best described with Band+BB model. The BB components are only significant for the time intervals $T_0+1.32$ to $T_0+3.84$\,s and $T_0+11.22$ to $T_0+11.86$\,s, and not appear in other time periods (See Figure \ref{Fig.BandBB}). Such evolution behavior of prompt spectra could be understood with the black hole (BH) central engine model.

In the GRB BH central engine model, two mechanisms are considered to power the relativistic jet: the neutrino-antineutrino annihilation mechanism, which liberates the gravitational energy from the accretion disk \citep{Popham1999, DiMatteo2002, Gu2006, Chen2007, Janiuk2007, Liu2007, Liu2017, Lei2009} and the Blandford-Znajek \citep[hereafter BZ]{Blandford1977} mechanism, which extracts the spin energy from the Kerr BH \citep{Lee2000, Li2000, Liu2015, Lei2013, Lei2017}.

The neutrino-antineutrino annihilation mechanism launches a thermal ``fireball'', leaving an imprint of the thermal component in GRB spectrum. The neutrino annihilation luminosity can be described with smooth power law with two breaks \citep{Lei2017}, i.e., 
\begin{eqnarray}
\dot{E}_{\nu \bar{\nu}}  \simeq  && \dot{E}_{\nu \bar{\nu}, \rm ign} \left[ \left(\frac{\dot{m}}{\dot{m}_{\rm ign}} \right)^{-\alpha_{\nu \bar{\nu}} } + \left(\frac{\dot{m}}{\dot{m}_{\rm ign} } \right)^{-\beta_{\nu \bar{\nu}} } \right]^{-1} \nonumber \\
&& \times   \left[1+(\frac{\dot{m}}{\dot{m}_{\rm trap} })^{\beta_{\nu \bar{\nu}} - \gamma_{\nu \bar{\nu}}  }\right]^{-1} ,
\label{eq_Evv}
\end{eqnarray}
where,
\begin{eqnarray}\label{Eq.Evv}
&&\left\lbrace
\begin{tabular}{l}
$\dot{E}_{\nu \bar{\nu}, \rm ign}=10^{(48.0+0.15 a_\bullet)}  \left(\frac{m_\bullet}{3} \right)^{\log(\dot{m}/\dot{m}_{\rm ign}) -3.3} {\rm erg \ s^{-1}},  $ \\
$\alpha_{\nu \bar{\nu}} = 4.7, \ \beta_{\nu \bar{\nu}} =  2.23, \  \gamma_{\nu \bar{\nu}} =0.3, $
\end{tabular} 
\right.\nonumber \\
&& \dot{m}_{\rm ign} = 0.07-0.063 a_\bullet,  \ \dot{m}_{\rm trap} = 6.0-4.0 a_\bullet^3,
\end{eqnarray}
where $\dot{m}_{\rm ign}$ and $\dot{m}_{\rm trap}$ are the igniting and trapping accretion rates, respectively. 

\begin{figure}[htbp!]
\plotone{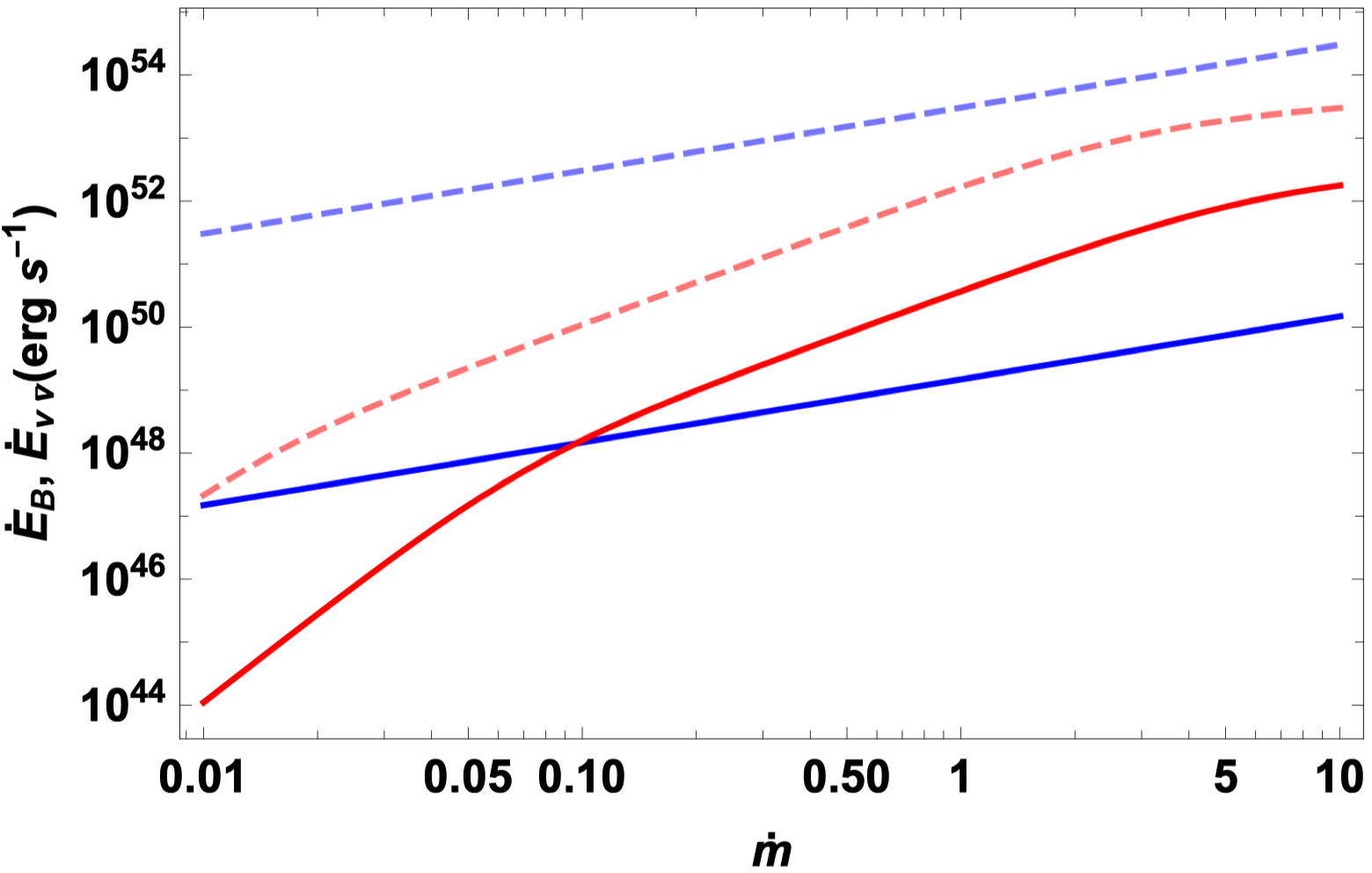}
\caption{The BZ power $\dot{E}_{\rm B} $ (blue lines) and neutrino annihilation power $\dot{E}_{\nu\bar{\nu}}$ (red lines) as a function of accretion rate $\dot{m}$ for different BH spin $a_\bullet =0.01$ (thick solid lines) and $0.9$ (dashed lines). We adopt the BH mass $m_\bullet=3$. Same results can also be found in \cite{Lei2017}.
\label{Fig.LvvLBZ}}
\end{figure}

However, the jet powered by the BZ mechanism is Poynting-flux-dominated. The BZ jet power from a BH with mass $M_{\bullet}$ and angular momentum $J_\bullet$ is \citep{Lee2000,Li2000,Wang2002,McKinney2005,Lei2005,Lei2011,Lei2013,Lei2017},

\begin{eqnarray}\label{Eq.EB}
 \dot{E}_{\rm B} & = & 9 \times 10^{53} a_\bullet^2 \dot{m}  X(a_\bullet) \ {\rm erg \ s^{-1}} 
\end{eqnarray}
where $X(a_\bullet)=F(a_\bullet)/(1+\sqrt{1-a_\bullet^2} )^2$,  and $F(a_\bullet)=[(1+q^2)/q^2][(q+1/q) \arctan q-1]$. Here $q= a_{\bullet} /(1+\sqrt{1-a^2_{\bullet}})$. The comparison between these two powers (the BZ power $\dot{E}_{\rm B} $ (blue lines) and neutrino annihilation power $\dot{E}_{\nu\bar{\nu}}$) is presented in Figure \ref{Fig.LvvLBZ}. In the following calculations, we adopt the BH mass $m_\bullet=3$ as a typical value.

One can see from Equations (\ref{Eq.Evv}) - (\ref{Eq.EB}) and Figure \ref{Fig.LvvLBZ}, both the neutrino-annihilation power and the BZ power depend on accretion rate $\dot{m}$ and BH spin $a_{\bullet}$ . However, the dependence on accretion rate is more serious for neutrino-annihilation power. Comparing these two mechanisms, we find that: (1) the magnetic power is much greater than the neutrino annihilation power for a moderate to high spin BH (see the dashed lines in Figure \ref{Fig.LvvLBZ}); (2) The neutrino annihilation power dominates over the BZ power for BHs with a very small spin at high accretion rates (see the thick solid lines in Figure \ref{Fig.LvvLBZ}); (3) compared with the magnetic power, $\dot{E}_{\nu \bar{\nu}}$ is much more sensitive to the mass accretion rate $\dot{m}$, especially at low accretion rates.

As a result, for low-spin BH, the jet can evolve from thermal significant via neutrino-annihilation mechanism to Poynting-flux dominated via BZ process if accretion rate drops, and from Poynting-flux dominated to thermal significant if reverse \citep{Lei2017}. 

\begin{figure}[htbp!]
\plotone{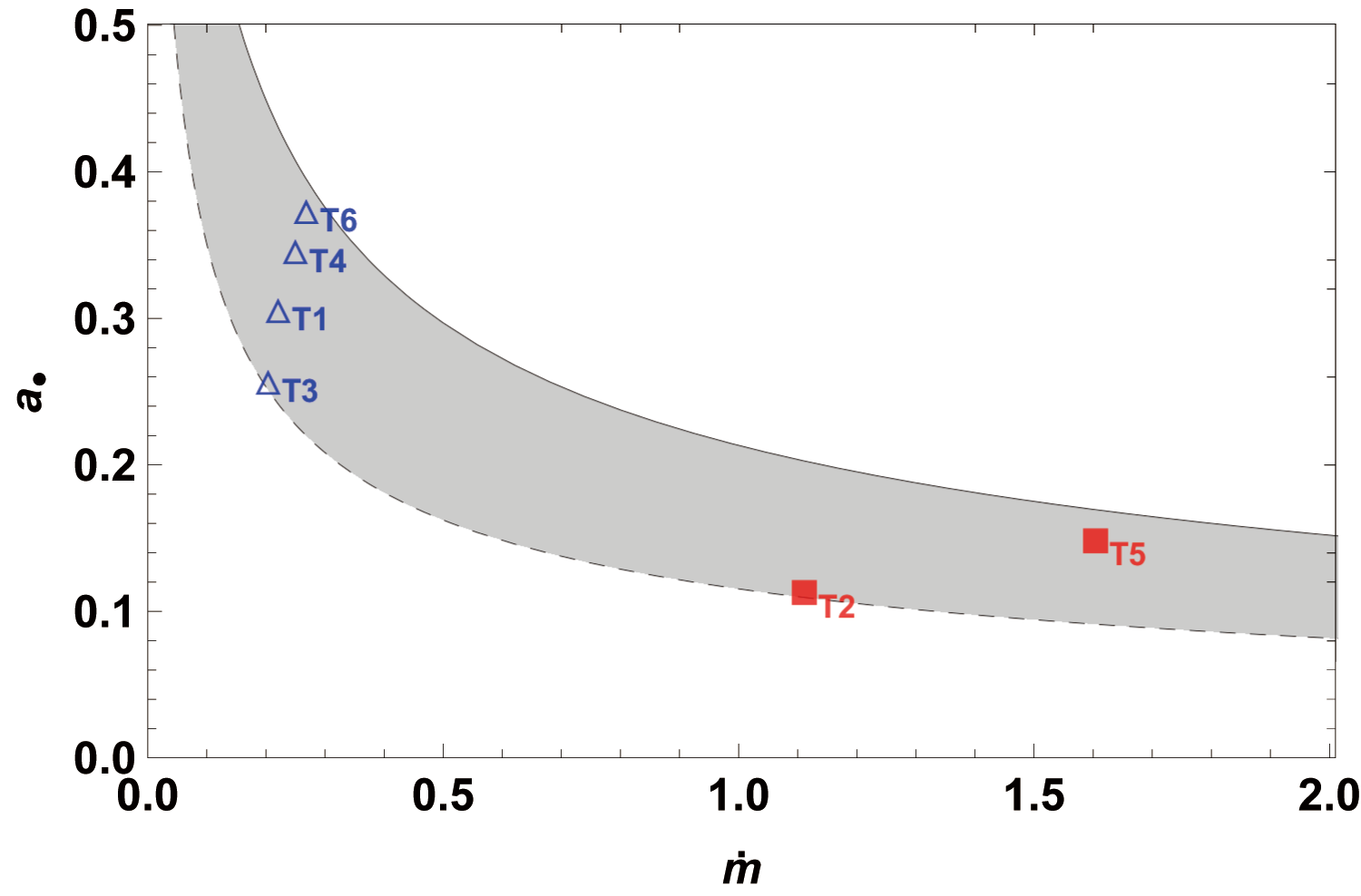}
\caption{The constraint on central engine parameters, i.e., the accretion rate $\dot{m}$ and BH spin $a_\bullet$, for GRB 200613A. The shaded region is obtained by equating $\dot{E}_{\rm B} \eta_{\gamma}/f_{\rm b} = 3.5\times 10^{52} \ {\rm erg \ s^{-1}}$ (the upper boundary, the solid black line) and $\dot{E}_{\rm B} \eta_{\gamma}/f_{\rm b} = 1\times 10^{52} \ {\rm erg \ s^{-1}}$ (lower boundary, the dashed black line). The blue triangles are the ones obtained by assuming the upper limit of BB component as $L_{\gamma,\rm iso}^{\rm BB} \leq L_{\gamma,\rm iso}^{\rm Band}/100 $. The true value might be left the blue triangles in the shaded region. The BH mass $m_\bullet=3$, efficiency $\eta_{\gamma} \simeq 0.45$ and beaming factore $f_{\rm b} = 0.09$ are adopted when producing this figure. 
\label{Fig.amdot}}
\end{figure}

Inspecting Figure \ref{Fig.BandBB}, GRB 200613A indeed shows significant spectral evolution in prompt emission. For convenience, we define the six time slides as T1 (-0.32 - 1.32 s), T2 (1.32 - 3.84), T3 (3.84 - 6.57 s), T4 (10.48 - 11.22 s), T5 (11.22 - 11.86 s) and T6 (11.86 - 13.84 s). Only in episodes T2 and T5, one can find significant thermal component, with BB component $L_{\gamma,\rm iso}^{\rm BB}=3.17_{-0.65}^{+0.57}\times 10^{51} \ \rm erg \ s^{-1} $ ($7.41_{-3.46}^{+2.41}\times 10^{51} \ \rm erg \ s^{-1} $) and Band component $L_{\gamma,\rm iso}^{\rm Band}=1.07_{-0.13}^{+0.17}\times 10^{52} \ \rm erg \ s^{-1} $ ($2.68_{-0.36}^{+0.41}\times 10^{52} \ \rm erg \ s^{-1} $) for T2 (T5). The central engine parameters can be constrained by relating the BB component to the neutrino annihilation power and the Band component to the BZ power, respectively, i.e., 
\begin{equation}
    \dot{E}_{\nu \bar{\nu}} \eta_{\gamma} = L_{\gamma,\rm iso}^{\rm BB} f_{\rm b} , \ \ \dot{E}_{\rm B} \eta_{\gamma} = L_{\gamma,\rm iso}^{\rm Band} f_{\rm b} ,
    \label{eq.BBvsBand}
\end{equation}
where $\eta_{\gamma} \equiv \frac{E_{\gamma,\rm iso} }{E_{\gamma,\rm iso} + E_{\rm K, iso} }$ is the efficiency of converting jet power to gamma-ray radiation and $f_{\rm b} \equiv 1-\cos\theta_{\rm j} \simeq \theta_{\rm j}^2/2$ is the beaming factor of the jet. We have $\eta_{\gamma} \simeq 0.45$ by substituting the observed gamma-ray isotropic energy $E_{\gamma,\rm iso}=1.68_{-0.06}^{+0.07}\times 10^{53}$ erg and model fit isotropic jet kinetic energy $E_{\rm K,iso}=2.04_{-1.50}^{+11.8}\times 10^{53}$ erg, and $f_{\rm b} = 0.09$ by adopting the model fit opening angle $\theta_{\rm j} \simeq 24$ degrees (see Table \ref{tb.AGmodel}). For episodes T1, T3, T4 and T6, the BB component can not be resolved from prompt spectra. We thus assume $L_{\gamma,\rm iso}^{\rm BB} \leq L_{\gamma,\rm iso}^{\rm Band}/100 $ as an upper limit to get a limit for the central engine parameters by using Equation (\ref{eq.BBvsBand}) for these time episodes. Since both BZ power and neutrino annihilation power are functions of the accretion rate and BH spin, one can constrain $\dot{m}$ and $a_\bullet$ based on the above assumptions, the results are presented in Figure \ref{Fig.amdot}.

From Figure \ref{Fig.amdot}, we can see that the BB component become significant (see episodes T2 and T5) for a BH central engine with low spin and high accretion rate. Therefore, the temporal behavior of GRB spectra in the prompt emission phase might provide meaningful clues to the central engine models. The evolution of BB component in prompt spectra may reflect the combination of these two mechanisms, and the evolution of central engine parameters (mainly the accretion rate) during the prompt emission in GRB 200613A.

We would also like to check the BH central engine model with other observations. The baryonic jet driven by neutrino annihilation model will produce a GRB via internal shock at $R_{\rm IS}$. The BZ mechanism launch a less baryonic jet, and convert the magnetic energy to gamma-rays by magnetic reconnection, e.g., ICMART (the Internal-Collision induced Magnetic Reconnection and Turbulence, \citet{Zhang2011ICMART}). In such case, many collisions are needed to trigger an ICMART event, so that one would have the gamma-ray emission radius $R_{\rm ICMART} \geq  R_{\rm IS}$. The estimated size of emission region $R_{\rm c} \simeq 5.7\times 10^{13}$ cm is a typical value for GRBs, we can not rule out any of the two models.

\cite{Lei2013} addressed the fundamental problem of baryon loading in GRB jets. For neutrino annihilation driven jet, the maximum available Lorentz factor $\Gamma_{\rm max} = \eta$ with \citep{Lei2013} 
\begin{eqnarray}
\eta  & \simeq   & 50 A^{-0.85} B^{1.35} C^{-0.22} \theta_{j,-1}^{-2} \alpha_{-1}^{-0.57} \epsilon_{-1}^{-1.7} \left(\frac{\xi}{2}\right)^{-0.32} \nonumber \\
& \times & \left(\frac{ R_{ms}}{2}\right)^{-5.12} \left(\frac{m}{3}\right)^{-0.6} \dot{m}^{0.55}, 
\label{eq:eta}
\end{eqnarray}
where $A, B, C$ are relativistic correction factors for a thin accretion disk around a Kerr BH, $\xi \equiv r/r_{\rm ms}$ is the disk radius in terms of the radius of the marginally stable orbit $r_{\rm ms}$, and $\alpha$ is the viscous parameter.
Here $\epsilon \simeq (1-E_{\rm ms})$ is the neutrino emission efficiency, and $E_{\rm ms}$ is the specific energy corresponding to $r_{\rm ms}$. The expressions for these parameters can be found in \citet{Lei2013}.  

For the BZ powered jet, one can define a parameter denoting the maximum available energy per baryon \citep{Lei2013} 
\begin{eqnarray}
\mu_{0} & \simeq & 1.5 \times 10^5 A^{-0.58} B^{0.83} F_{\rm p,-1}^{0.5} \theta_{\rm j,-1}^{-1}  \theta_{\rm B,-2}  \alpha_{-1}^{-0.38} \epsilon_{-1}^{-0.83}    \nonumber \\
& & a_\bullet^2   X(a_\bullet)  \left(\frac{m}{3}\right)^{-0.55} \dot{m}_{-1}^{0.17} r_{z,11}^{-0.5},
\label{eq:mu}
\end{eqnarray}
where $F_{\rm p}$ is the fraction of protons, $r_{\rm z}$ is the distance from the BH in the jet direction. Protons with an ejected direction larger than $\theta_{\rm B}$ with respect to the field lines would be blocked because of the existence of a strong magnetic field. The BZ jet will reach a terminating Lorentz factor $\Gamma$ that satisfies $\Gamma_{\rm min} < \Gamma < \Gamma_{\rm max}$ with the explicit value depending on the detailed dissipation process. We take $\Gamma_{\rm min}=\mu_{0}^{1/3}$ and $\Gamma_{\rm max} =  \mu_0$ for efficient magnetic acceleration. 

\begin{figure}[htbp!]
\plotone{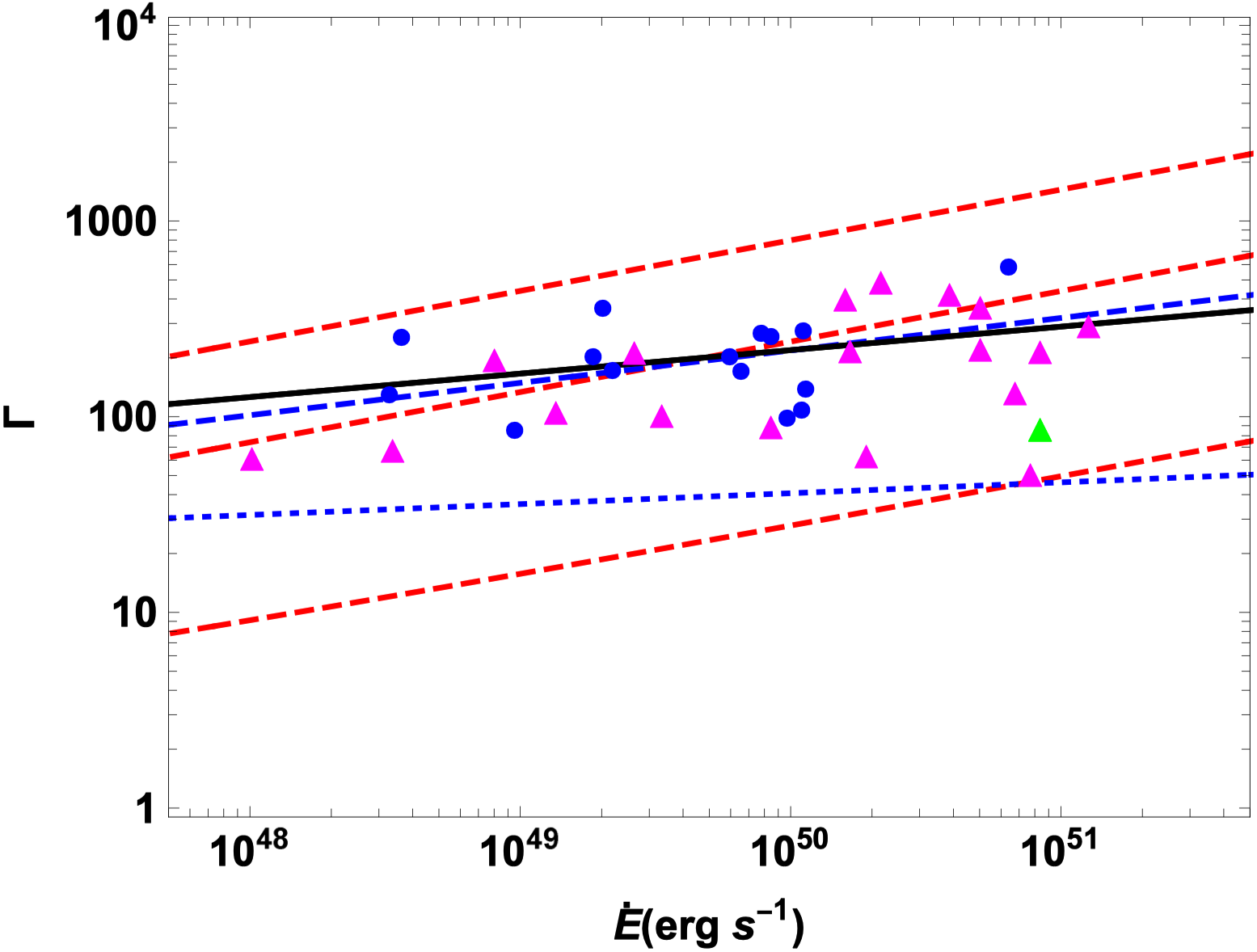}
\caption{A comparison between the observations and the predictions from the neutrino annihilation mechanism (red dashed lines) and the BZ mechanism (blue dashed lines), reproduced from \citet{Yi2017}. The blue dots indicate GRBs with the beaming corrected luminosity, whereas magenta triangles are the ones with only lower limit. The red dashed lines show $\eta(\dot{m}) - \dot{E}_{\nu\bar{\nu}} (\dot{m})$ relation for $\theta_{\rm j}$ = 0.022 (top), 0.04(middle) and 0.12 (bottom), respectively. The blue dashed line shows the $\Gamma_{\rm max} (\dot{m})$ - $\dot{E}_{\rm B} (\dot{m})$ relation. The $\Gamma_{\rm min} (\dot{m})$ - $\dot{E}_{\rm B} (\dot{m})$ is plotted with the blue dotted line. The best-fit to the data $\Gamma_0 \propto L_{\rm tot}^{0.14}$ are shown with a sold line \citep{Yi2017}. The green triangles indicate the GRB 200613A with lower limit.
\label{Fig.GammaL}}
\end{figure}

It is found that a magnetically dominated jet can be much cleaner and is more consistent with the requirement of large Lorentz factors in GRBs. For GRB 200316A, we only obtain a minimum Lorentz factor $\Gamma_{\rm min} \leq 89$, and the neutrino annihilation model is still possible jet launching model although it is typically ``dirtier'' \citep{Lei2013}. To show this, in Figure \ref{Fig.GammaL} we reproduce the $\Gamma$ - $\dot{E}$ relation (including observational data and model predictions) in \citet[Figure 3 therein]{Yi2017}, and the observational result of GRB 200613A with green triangle. The beaming corrected total jet luminosity is detemined by $L_{\rm j} = (E_{\gamma,\rm iso}+E_{\rm K, iso}) f_{\rm b}/T_{90}$, which is $\sim 8.4\times 10^{50} \ {\rm erg \ s^{-1}}$ for GRB 200613A. From Figure \ref{Fig.GammaL}, one finds that both neutrino annihilation model and BZ mechanism can work for GRB 200613A. 

\subsection{Closure Relations}\label{dis.CR}

The multi-band light curve of GRB\,200613A from 0.5 days to 19 days is presented in Figure \ref{Fig.lc}. The optical flux displays a smooth decay ($\alpha_{opt}\sim 0.75$) with no discernible break, while the X-ray flux follows a relatively steep decay index ($\alpha_{X_1}\sim 1.04$) within a week. This chromatic evolution between X-ray and optical may indicate that the X-ray energy band and optical band are in different spectral regimes. A natural explanation is that the cooling break ($\nu_c$) is located between the optical ($\nu_{opt}$) and X-ray bands ($\nu_{X}$). According to closure relation \citep[e.g.][]{2004IJMPA..19.2385Z, 2013NewAR..57..141G}, different circumburst density proﬁle predict different temporal and spectral indices. The expected decline rate of X-ray is $\alpha_X = (3p-2)/4$ for both the ISM case and the wind case, and one can derive the electron energy distribution index $p=(4\alpha_X+2)/3=2.05$. However, the expected r-band decline rate is $\alpha_{r,ISM} = 3(p-1)/4=0.78$ for ISM case and  $\alpha_{r,wind} = (3p-1)/4=1.29$ for wind case. Therefore, the ISM-like circumburst medium is more consistent with observational results. Time-averaged spectrum of X-ray gives the photon index\footnote{\url{https://www.swift.ac.uk/xrt_spectra/00021003/}} of $\Gamma = 1.91_{-0.14}^{+0.15}$, i.e. $\beta_X = 0.91_{-0.14}^{+0.15}$ \citep{2009MNRAS.397.1177E}. The expected spectral index $\beta_{X,exp} = p/2 = 1.03$ $\sim \beta_{X,obs}=0.91\pm0.15$, which is consistent with the fitting results of the X-ray SED.

The X-ray flux begins to decay more steeply than before ($\alpha_{X_2}\sim 1.5$) after a week. However, this effect is not observed in the optical band. We speculate that the emission of the host galaxy may counteract the break effect in the optical, thus rendering the break effect in the optical is undetectable at late times. The most possible explanation for the X-ray break is the jet-break effect. According to the theoretical prediction, the change of temporal index is $\Delta \alpha_{jb} = 0.75$ \citep[e.g.][]{1998ApJ...503..314P, 1999MNRAS.306L..39M, 2013NewAR..57..141G}, which is consistent with the temporal fitting result when considering the large uncertainty.

\subsection{Comparison with GRB host galaxy samples}

LGRBs are believed to originate from massive stellar explosions, which have relatively short lifetimes. Consequently, such events tend to occur in star-forming galaxies or star-forming regions of the galaxy. In a recent study by \cite{2022A&A...666A..14S}, a sample of 44 long GRB host galaxies observed with the HST/WFC3 in the IR band at redshift range from 1 to 3.1 was examined. The study revealed no systematic offset in physical properties between LGRBs host galaxies and star-forming galaxies. However, when the samples were grouped by redshift, it was found that the host galaxies in the redshift range of $1<z<2$ exhibited higher stellar mass and star formation density than expected, while host galaxies in the range of $2<z<3.1$ were more consistent with star-forming galaxies in the field.

On the other hand, short-duration gamma-ray bursts (SGRBs) are thought to originate from the coalescence of compact binary systems, which have a long delay time between binary formation and merger. Studies of some pulsars have shown that such ultra-compact objects receive a significant kick velocity at birth \citep{1995ApJ...452..819L, 1996ApJ...466L..35L, kaspi1996evidence}. Recent numerical modeling studies, however, suggest that different natal-kick mechanisms can result in different mean natal-kick values for NS binaries. For instance, the collapse-asymmetry mechanism may result in lower kicks for NS binaries compared to other mechanisms, potentially biasing the overall mean natal-kick value for this population \citep[e.g.][]{2018arXiv181210065B, 2020A&A...639A..41B}. Consequently, SGRBs can occur in various locations within the host galaxy and are not limited to star-forming regions. \cite{2022ApJ...940...56F} demonstrated that the host-normalized offset for SGRBs is 2.5 times larger than that observed for LGRBs. Furthermore, due to their long formation timescales, SGRBs can also occur in different types of galaxies. Recent research has indicated that the majority of SGRBs are associated with star-forming galaxies \citep{2010ApJ...725.1202L, 2013ApJ...769...56F, 2014ARA&A..52...43B}, and their hosts exhibit higher stellar mass and lower specific star formation rates \citep{2022ApJ...940...57N}.

\begin{figure}[htbp!]
\plotone{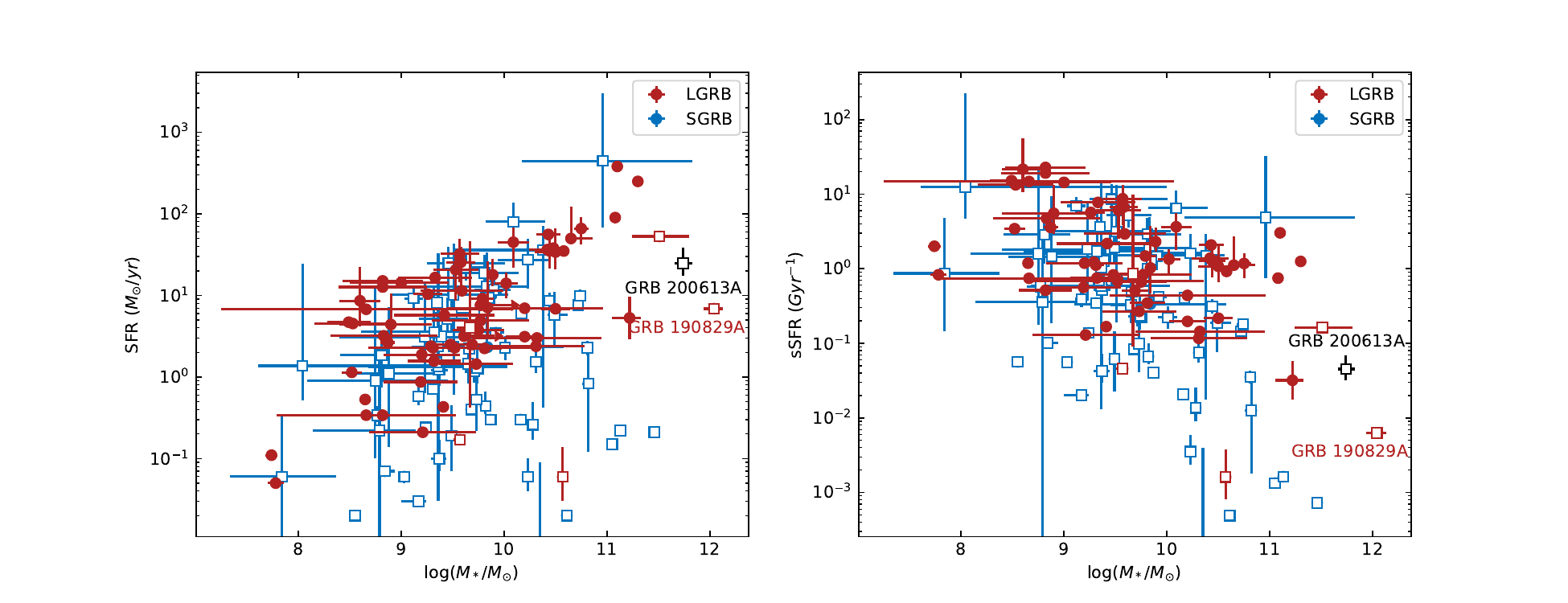}
\caption{The stellar mass of the host galaxies versus the star formation rate (left panel) and specific star formation rate (right panel) of our sample. Open squares represent parameters derived using \texttt{Prospector}, while solid circles represent those derived in other ways (e.g., via other numerical models or mass-to-light relations). The host galaxy of GRB\,200613A is represented by a black open square. Notably, the most massive host galaxy in our sample is GRB\,190829A.  \label{Fig.host_compare}}
\end{figure}

To conduct a comparative analysis of the host galaxy of GRB200613A with a representative sample of GRB hosts, we assembled a collection of LGRB and SGRB host galaxies from \cite{2009ApJ...691..182S}, \cite{2022A&A...666A..14S}, \cite{2022ApJ...940...57N}, \cite{2022JApA...43...82G}, \cite{2021NatAs...5..917A}, \cite{2023arXiv230814197B} and \cite{2023NatAs...7..976L}, which consists of 89 LGRB hosts and 72 SGRB hosts. 
As depicted in Figure \ref{Fig.host_compare}, the host galaxy of GRB\,200613A is identified as the second most massive galaxy within our sample, with the exception of the host galaxy of GRB\,190829A as reported by \cite{2022JApA...43...82G}. The star formation rate of GRB\,200613A's host galaxy falls within a moderate range when compared to other LGRB samples.

In Figure \ref{Fig.host_compare}, six LGRBs' host derived from \texttt{Prospector} deviate from the other LGRBs' host, generally located in the lower right corner of the mass-SFR diagram. These LGRBs' host samples are from different literature sources and employ diverse calculation methods, leading to systematic biases in the derived stellar masses and SFRs. As emphasized in \cite{2022A&A...666A..14S}, different methods can yield stellar mass that differ by nearly an order of magnitude. Underlying assumptions within galaxy SED models, such as the dust attenuation, IMF, and SFH, can introduce discrepancies in derived stellar masses and SFRs. In the MAGPHYS model \citep{2008MNRAS.388.1595D}, for example, using a Salpeter IMF \citep{1955ApJ...121..161S} instead of the Chabrier IMF \citep{2003PASP..115..763C} would result in stellar mass estimates that are 40\% larger. Our [OII]-derived SFR are approximately 50\% higher than those derived from SED fitting, although they are still within error margins. This observation aligns with \cite{2022ApJ...940...57N}, who found most emission-line SFR estimates exceed those from SED fitting. \cite{2019ApJ...877..140L} demonstrated that contributions from "older" stellar populations and AGNs can lead to lower SFR compared to those derived from $L_{\rm UV+IR}$. Therefore, implementing a consistent approach for all LGRBs' host in estimating stellar masses and SFRs would undoubtedly enhance result comparability. However, this lies beyond the scope of the present study. Even after considering these systematic biases, the host galaxy masses remain comparatively large within the sample.

\section{Summary}
We present our optical photometry observation of GRB\,200613A afterglow using OAJ T80 telescope, Xinglong 2.16m telescope, NOT 2.56 m telescope and CAHA 2.2m telescope, as well as spectroscopy of host galaxy using 10.4m-GTC telescope. We conducted a comprehensive analysis of the multi-band radiation of GRB\,200613A, which included gamma-ray emission in prompt phase, X-ray, and optical afterglow. We also utilized the photometric data obtained from SDSS, Legacy Survey, and WISE to analyze the properties of its host galaxy. The main results of our study are summarized as follows: 
\begin{itemize}
    \item [1.] We obtain the host galaxy spectrum of GRB\,200613A using 10.4m-GTC telescope 218 days after the burst. According to prominent emission lines [OII] 3727/3729 doublet, we get a redshift of $z = 1.2277 \pm 0.0011$ for GRB\,200613A, and a similar redshift of $z = 1.2262 \pm 0.0009$ for the likely intervening system by the absorption line.
    \item [2.] The combined evidence of long duration $T_{90}$, location on the lower-right edge of the Amati relation diagram, and significant spectral lag strongly suggests that GRB\,200613A is a typical collapse burst (Type II burst).
    \item [3.] The time-resolved spectral analysis of the prompt gamma-ray emission reveals that the Band component dominates the entire time period, with an additional blackbody component appearing in the early phase and at the time of the flux peak. We speculate that the main Band component comes from the Blandford-Znajek mechanism, and the additional BB component comes from the neutrino annihilation mechanism. Compared with observed data, we find that the BB component become significant (see episodes T2 and T5) for a BH central engine with low spin and high accretion rate. Therefore, the temporal behavior of GRB spectra in the prompt emission phase provide meaningful trace of the central engine.
    \item [4.] The multi-band afterglow shows a power-law decay, with signs of a jet break seen in the X-ray band seven days after the burst. This afterglow can be explained by the model of a jet colliding with the interstellar medium (ISM), with the physical parameters: $E_{K,iso} = (2.04^{+11.8}_{-1.50})\times 10^{53}$ erg, $\Gamma_0=354^{+578}_{-217}$, $p=2.09^{+0.02}_{-0.03}$, $n_{18}=(2.04^{+9.71}_{-1.87})\times 10^{2}$ cm$^{-3}$, $\theta_j=24.0^{+6.50}_{-5.54}$ degree, $\epsilon_e=1.66^{+4.09}_{-1.39})\times 10^{-1}$ and $\epsilon_B=(7.76^{+48.5}_{-5.9})\times 10^{-6}$. Due to the lack of early afterglow observation, the initial Lorentz factor, $\Gamma_0$, cannot be constrained. We can only derive a lower bound of $\Gamma_{0,min} \sim 89$ based on the minimum variability scales of the prompt emission.
    \item [5.] We collected the magnitudes of the host galaxy from the SDSS, Legacy Survey, and WISE sky surveys and modeled its spectral energy distribution (SED) using the \texttt{Prospector} Python package. The results suggest that the host galaxy is a massive galaxy with a stellar mass of $\log(M_\ast / M_\odot)=11.75^{+0.10}_{-0.09}$ and a moderate star formation rate of $\mbox{SFR}=22.58^{+13.63}_{-7.22} M_{\odot}$/yr in our GRB host galaxies sample. We additionally employed the [OII] emission line to independently estimate SFR of the host galaxy and the results ($\sim 46.4 M_{\odot}$yr$^{-1}$ if $\langle A_V \rangle = 0.53$ and $\sim 34.2 M_{\odot}$ yr$^{-1}$ if $A_V = 0.28$) were basically consistent with the value derived from \texttt{Prospector}.
\end{itemize}

\begin{acknowledgments}

We acknowledge the support of the staff of the Xinglong 2.16-m telescope. This work was also partially supported by the Open Project Program of the Key Laboratory of Optical Astronomy, National Astronomical Observatories, Chinese Academy of Sciences.
This publication is partly based on observations made with the Gran Telescopio Canarias (GTC), within programs GTC06-20ADDT (PI: Antonio de Ugarte Postigo) and GTCMULTIPLE2G-20B (PI: Antonio de Ugarte Postigo).
This publication is partly based on observations made with the Calar Alto Observatory (CAHA), within program F20-2.2-024 (PI: Antonio de Ugarte Postigo)
This publication is partly based on observations made with the Javalambre Observatory (OAJ), within program 1900170 (PI: David Alexander Kann). 
The data presented here were obtained in part with ALFOSC, which is provided by the Instituto de Astrofisica de Andalucia (IAA) under a joint agreement with the University of Copenhagen and NOT. This research has made use of the Spanish Virtual Observatory (http://svo.cab.inta-csic.es) supported by the MINECO/FEDER through grant AyA2017-84089.7. 
This work is supported by the National Key R\&D Program of China (Nos. 2020YFC2201400), the National Natural Science Foundation of China under grants U2038107 and U1931203. W. H. Lei acknowledges support by the science research grants from the China Manned Space Project with NO.CMS-CSST-2021-B11.
This work made use of data supplied by the UK Swift Science Data Centre at the University of Leicester.
This publication makes use of data products from the Wide-field Infrared Survey Explorer, which is a joint project of the University of California, Los Angeles, and the Jet Propulsion Laboratory/California Institute of Technology, funded by the National Aeronautics and Space Administration.

\end{acknowledgments}

\vspace{5mm}
\facilities{XingLong 2.16m(BFOSC), 10.4m GTC (OSIRIS), 2.2m CAHA(CAFOS), JAST80 (T80Cam), NOT(ALFOSC), Swift(XRT and UVOT), Fermi(LAT and GBM)}

\software{Astropy \citep{2013A&A...558A..33A,2018AJ....156..123A}, 
          Source Extractor \citep{1996A&AS..117..393B}, 
          emcee \citep{2013PASP..125..306F}, 
          PyRAF \citep{1986SPIE..627..733T, 1993ASPC...52..173T}, 
          Prospector \citep{2021ApJS..254...22J},
          3ML \citep{3ml}
          }

\bibliography{main}{}
\bibliographystyle{aasjournal}

\appendix
\restartappendixnumbering

\section{Time-resolved spectral analysis results}

\startlongtable
\begin{deluxetable*}{llclllll}
\tablecaption{Time-resolved spectral analysis results of prompt emission}
\label{tb.prompt_sed}
\tablewidth{0pt}
\tablehead{
\colhead{$t_{start}$ (s)} & \colhead{$t_{stop}$ (s)} & \colhead{Model} & \colhead{BIC} & \colhead{$\alpha$} & \colhead{$\beta$} & \colhead{$E_p$} & \colhead{$kT$} 
}
\startdata
        -0.32  & 1.32  & Band & 2693.99 & $-0.51^{+0.11}_{-0.1}$ & $-2.78^{+0.37}_{-0.43}$ & $623.87^{+93.63}_{-76.97}$ & \nodata \\ 
        1.32  & 3.84  & Band & 3342.7 & $-0.16^{+0.12}_{-0.11}$ & $-2.39^{+0.14}_{-0.19}$ & $152.91^{+12.81}_{-11.42}$ & \nodata \\ 
        3.84  & 6.57  & Band & 3483.79 & $-0.64^{+0.08}_{-0.07}$ & $-2.78^{+0.12}_{-0.14}$ & $75.22^{+3.24}_{-3.35}$ & \nodata \\ 
        6.57  & 9.13  & Band & 3287.12 & $-1.09^{+0.09}_{-0.1}$ & $-2.56^{+0.1}_{-0.13}$ & $51.89^{+2.43}_{-1.32}$ & \nodata \\ 
        9.13  & 9.80  & Band & 1407.77 & $-1.43^{+0.06}_{-0.05}$ & $-2.88^{+0.37}_{-0.42}$ & $145.52^{+22.65}_{-19.69}$ & \nodata \\ 
        9.80  & 10.48  & Band & 1468.66 & $-1.3^{+0.06}_{-0.06}$ & $-3.04^{+0.36}_{-0.4}$ & $163.39^{+19.86}_{-16.58}$ & \nodata \\ 
        10.48  & 11.22  & Band & 1630.04 & $-1.17^{+0.05}_{-0.04}$ & $-3.2^{+0.33}_{-0.37}$ & $238.51^{+23.69}_{-20.74}$ & \nodata \\ 
        11.22  & 11.86  & Band & 1516.66 & $-0.85^{+0.05}_{-0.05}$ & $-3.17^{+0.29}_{-0.35}$ & $200.5^{+12.08}_{-11.27}$ & \nodata \\ 
        11.86  & 13.84  & Band & 3186.8 & $-0.7^{+0.03}_{-0.03}$ & $-3.48^{+0.28}_{-0.31}$ & $180.69^{+5.09}_{-5.16}$ & \nodata \\ 
        13.84  & 14.82  & Band & 1997.08 & $-0.83^{+0.05}_{-0.05}$ & $-3.42^{+0.3}_{-0.34}$ & $123.18^{+5.6}_{-5.4}$ & \nodata \\ 
        14.82  & 15.64  & Band & 1685.13 & $-0.8^{+0.07}_{-0.07}$ & $-3.34^{+0.27}_{-0.31}$ & $96.85^{+4.66}_{-4.23}$ & \nodata \\ 
        15.64  & 16.13  & Band & 1062.99 & $-0.87^{+0.13}_{-0.11}$ & $-3.12^{+0.31}_{-0.36}$ & $86.07^{+7}_{-6.57}$ & \nodata \\ 
        16.13  & 18.81  & Band & 3275.07 & $-1.08^{+0.05}_{-0.05}$ & $-3.26^{+0.26}_{-0.32}$ & $96.35^{+4.45}_{-4.11}$ & \nodata \\ 
        18.81  & 21.15  & Band & 3096.59 & $-1.06^{+0.06}_{-0.06}$ & $-3.2^{+0.27}_{-0.32}$ & $99.53^{+5.54}_{-4.95}$ & \nodata \\ 
        21.15  & 22.38  & Band & 2050.54 & $-0.94^{+0.14}_{-0.13}$ & $-2.89^{+0.28}_{-0.34}$ & $76.6^{+7.29}_{-6.66}$ & \nodata \\ 
        22.38  & 27.77  & Band & 4313.25 & $-0.97^{+0.1}_{-0.1}$ & $-2.93^{+0.17}_{-0.22}$ & $52.44^{+2.24}_{-1.61}$ & \nodata \\ 
        27.77  & 30.91  & Band & 3361.75 & $-1.3^{+0.11}_{-0.1}$ & $-3.35^{+0.29}_{-0.34}$ & $51^{+1.27}_{-0.72}$ & \nodata \\ 
        30.91  & 33.62  & Band & 3098.74 & $-1.39^{+0.11}_{-0.08}$ & $-3.21^{+0.3}_{-0.35}$ & $51.34^{+1.74}_{-0.96}$ & \nodata \\ 
        33.62  & 37.96  & Band & 3791.64 & $-1.45^{+0.06}_{-0.04}$ & $-2.84^{+0.31}_{-0.39}$ & $51.75^{+2.38}_{-1.27}$ & \nodata \\ 
        37.96  & 47.74  & Band & 4953.63 & $-1.44^{+0.07}_{-0.04}$ & $-3.16^{+0.33}_{-0.37}$ & $51.57^{+2.14}_{-1.13}$ & \nodata \\ 
        47.74  & 58.90  & Band & 5084.68 & $-1.39^{+0.13}_{-0.08}$ & $-2.88^{+0.38}_{-0.43}$ & $54.18^{+5.67}_{-3.02}$ & \nodata \\ 
        -0.32  & 1.32  & BB & 2690 & \nodata & \nodata & \nodata & $82.12^{+4.51}_{-4.19}$ \\ 
        1.32  & 3.84  & BB & 3461.25 & \nodata & \nodata & \nodata & $33.06^{+0.81}_{-0.79}$ \\ 
        3.84  & 6.57  & BB & 4144.9 & \nodata & \nodata & \nodata & $16.85^{+0.07}_{-0.07}$ \\ 
        6.57  & 9.13  & BB & 4293.82 & \nodata & \nodata & \nodata & $15.44^{+0.08}_{-0.08}$ \\ 
        9.13  & 9.80  & BB & 1819.97 & \nodata & \nodata & \nodata & $17^{+0.17}_{-0.16}$ \\ 
        9.80  & 10.48  & BB & 1997.6 & \nodata & \nodata & \nodata & $18.46^{+0.39}_{-0.28}$ \\ 
        10.48  & 11.22  & BB & 2498.91 & \nodata & \nodata & \nodata & $24.14^{+0.63}_{-0.61}$ \\ 
        11.22  & 11.86  & BB & 2185.88 & \nodata & \nodata & \nodata & $29.53^{+0.65}_{-0.62}$ \\ 
        11.86  & 13.84  & BB & 5330.19 & \nodata & \nodata & \nodata & $29.02^{+0.31}_{-0.31}$ \\ 
        13.84  & 14.82  & BB & 2753.49 & \nodata & \nodata & \nodata & $20.46^{+0.26}_{-0.19}$ \\ 
        14.82  & 15.64  & BB & 2118.04 & \nodata & \nodata & \nodata & $19.22^{+0.12}_{-0.11}$ \\ 
        15.64  & 16.13  & BB & 1259.99 & \nodata & \nodata & \nodata & $18.06^{+0.16}_{-0.16}$ \\ 
        16.13  & 18.81  & BB & 4293.41 & \nodata & \nodata & \nodata & $16.74^{+0.08}_{-0.08}$ \\ 
        18.81  & 21.15  & BB & 3745.19 & \nodata & \nodata & \nodata & $16.3^{+0.26}_{-0.2}$ \\ 
        21.15  & 22.38  & BB & 2191.87 & \nodata & \nodata & \nodata & $15.04^{+0.18}_{-0.15}$ \\ 
        22.38  & 27.77  & BB & 4771.6 & \nodata & \nodata & \nodata & $13.4^{+0.07}_{-0.07}$ \\ 
        27.77  & 30.91  & BB & 3574.2 & \nodata & \nodata & \nodata & $12.52^{+0.1}_{-0.11}$ \\ 
        30.91  & 33.62  & BB & 3155.48 & \nodata & \nodata & \nodata & $11.22^{+0.16}_{-0.15}$ \\ 
        33.62  & 37.96  & BB & 3845.04 & \nodata & \nodata & \nodata & $10.33^{+0.16}_{-0.15}$ \\ 
        37.96  & 47.74  & BB & 4969.07 & \nodata & \nodata & \nodata & $8.67^{+0.21}_{-0.18}$ \\ 
        47.74  & 58.90  & BB & 5074.36 & \nodata & \nodata & \nodata & $7.85^{+0.64}_{-0.44}$ \\ 
        -0.32  & 1.32  & BB+PL & 2587.96 & $-1.43^{+0.05}_{-0.06}$ & \nodata & \nodata & $92.61^{+7.48}_{-6.89}$ \\ 
        1.32  & 3.84  & BB+PL & 3245.16 & $-1.48^{+0.03}_{-0.03}$ & \nodata & \nodata & $33.07^{+1.22}_{-1.18}$ \\ 
        3.84  & 6.57  & BB+PL & 3434.79 & $-1.83^{+0.03}_{-0.03}$ & \nodata & \nodata & $17.55^{+0.57}_{-0.56}$ \\ 
        6.57  & 9.13  & BB+PL & 3247.3 & $-1.94^{+0.03}_{-0.03}$ & \nodata & \nodata & $12.03^{+0.4}_{-0.33}$ \\ 
        9.13  & 9.80  & BB+PL & 1331.82 & $-1.85^{+0.04}_{-0.05}$ & \nodata & \nodata & $27.74^{+4.64}_{-4.28}$ \\ 
        9.80  & 10.48  & BB+PL & 1389.46 & $-1.84^{+0.04}_{-0.05}$ & \nodata & \nodata & $30.82^{+2.64}_{-2.49}$ \\ 
        10.48  & 11.22  & BB+PL & 1577.85 & $-1.78^{+0.03}_{-0.03}$ & \nodata & \nodata & $40.07^{+2.37}_{-2.26}$ \\ 
        11.22  & 11.86  & BB+PL & 1435.15 & $-1.74^{+0.03}_{-0.04}$ & \nodata & \nodata & $38.45^{+1.49}_{-1.46}$ \\ 
        11.86  & 13.84  & BB+PL & 3206.63 & $-1.69^{+0.02}_{-0.02}$ & \nodata & \nodata & $35.03^{+0.7}_{-0.67}$ \\ 
        13.84  & 14.82  & BB+PL & 1952.73 & $-1.81^{+0.03}_{-0.03}$ & \nodata & \nodata & $25.96^{+0.97}_{-0.94}$ \\ 
        14.82  & 15.64  & BB+PL & 1614.66 & $-1.93^{+0.05}_{-0.05}$ & \nodata & \nodata & $22.99^{+0.91}_{-0.87}$ \\ 
        15.64  & 16.13  & BB+PL & 999.76 & $-1.89^{+0.06}_{-0.06}$ & \nodata & \nodata & $19.48^{+1.45}_{-1.37}$ \\ 
        16.13  & 18.81  & BB+PL & 3247.84 & $-1.94^{+0.03}_{-0.03}$ & \nodata & \nodata & $22.81^{+0.94}_{-0.92}$ \\ 
        18.81  & 21.15  & BB+PL & 3028.63 & $-1.93^{+0.04}_{-0.04}$ & \nodata & \nodata & $23.48^{+1.11}_{-1.06}$ \\ 
        21.15  & 22.38  & BB+PL & 1977.08 & $-1.9^{+0.06}_{-0.07}$ & \nodata & \nodata & $17.3^{+1.44}_{-1.31}$ \\ 
        22.38  & 27.77  & BB+PL & 4255.62 & $-1.97^{+0.04}_{-0.04}$ & \nodata & \nodata & $12.37^{+0.65}_{-0.6}$ \\ 
        27.77  & 30.91  & BB+PL & 3297.07 & $-2.12^{+0.06}_{-0.07}$ & \nodata & \nodata & $11.4^{+1.2}_{-0.92}$ \\ 
        30.91  & 33.62  & BB+PL & 2873.76 & $-2.18^{+0.08}_{-0.1}$ & \nodata & \nodata & $10.8^{+1.93}_{-1.31}$ \\ 
        33.62  & 37.96  & BB+PL & 3376.19 & $-2.11^{+0.06}_{-0.06}$ & \nodata & \nodata & $85.19^{+76.57}_{-66.98}$ \\ 
        37.96  & 47.74  & BB+PL & 4479.78 & $-2.25^{+0.07}_{-0.08}$ & \nodata & \nodata & $80.51^{+77.14}_{-61.83}$ \\ 
        47.74  & 58.90  & BB+PL & 4663.33 & $-2.18^{+0.11}_{-0.13}$ & \nodata & \nodata & $90.65^{+72.12}_{-68.01}$ \\ 
        -0.32 & 1.32 & Band+BB & 2401.55 & $-0.72^{+0.04}_{-0.1}$ & $-2.69^{+0.35}_{-0.43}$ & $815.04^{+182.82}_{-132.32}$ & $96.99^{+52.29}_{-33.18}$ \\ 
        1.32 & 3.84 & Band+BB & 3242.65 & $-1.01^{+0.19}_{-0.17}$ & $-2.65^{+0.43}_{-0.49}$ & $445.66^{+228.51}_{-132.61}$ & $29.22^{+1.97}_{-2.0}$ \\ 
        3.84 & 6.57 & Band+BB & 3167.61 & $-0.71^{+0.03}_{-0.05}$ & $-2.86^{+0.15}_{-0.14}$ & $77.73^{+2.87}_{-2.38}$ & $83.12^{+66.27}_{-67.28}$ \\ 
        6.57 & 9.13 & Band+BB & 2953.73 & $-1.1^{+0.1}_{-0.1}$ & $-2.56^{+0.11}_{-0.14}$ & $51.77^{+2.33}_{-1.21}$ & $69.41^{+79.75}_{-58.8}$ \\ 
        9.13 & 9.80 & Band+BB & 1087.4 & $-1.42^{+0.07}_{-0.05}$ & $-2.92^{+0.38}_{-0.43}$ & $142.98^{+20.85}_{-19.56}$ & $76.52^{+71.54}_{-55.09}$ \\ 
        9.80 & 10.48 & Band+BB & 1145.82 & $-1.3^{+0.06}_{-0.06}$ & $-3.06^{+0.37}_{-0.39}$ & $163.42^{+20.56}_{-17.93}$ & $85.24^{+71.91}_{-56.07}$ \\ 
        10.48 & 11.22 & Band+BB & 1220.39 & $-1.17^{+0.05}_{-0.05}$ & $-3.18^{+0.31}_{-0.37}$ & $237.03^{+22.45}_{-22.42}$ & $92.94^{+67.45}_{-59.67}$ \\ 
        11.22 & 11.86 & Band+BB & 1300.36 & $-1.13^{+0.2}_{-0.17}$ & $-2.91^{+0.39}_{-0.48}$ & $205.08^{+46.98}_{-44.85}$ & $42.65^{+7.12}_{-4.66}$ \\ 
        11.86 & 13.84 & Band+BB & 2861.4 & $-0.71^{+0.03}_{-0.03}$ & $-3.51^{+0.33}_{-0.31}$ & $182.16^{+5.86}_{-5.65}$ & $53.39^{+73.36}_{-22.62}$ \\ 
        13.84 & 14.82 & Band+BB & 1611.61 & $-0.82^{+0.06}_{-0.05}$ & $-3.4^{+0.31}_{-0.35}$ & $122.57^{+5.45}_{-5.12}$ & $79.84^{+75.9}_{-51.58}$ \\ 
        14.82 & 15.64 & Band+BB & 1390.04 & $-0.82^{+0.08}_{-0.09}$ & $-3.3^{+0.29}_{-0.34}$ & $97.14^{+5.27}_{-5.32}$ & $35.37^{+104.73}_{-15.27}$ \\ 
        15.64 & 16.13 & Band+BB & 688.36 & $-0.88^{+0.12}_{-0.11}$ & $-3.16^{+0.31}_{-0.31}$ & $86.41^{+7.11}_{-6.34}$ & $79.09^{+79.24}_{-59.24}$ \\ 
        16.13 & 18.81 & Band+BB & 2890.32 & $-1.08^{+0.05}_{-0.05}$ & $-3.24^{+0.25}_{-0.31}$ & $95.55^{+4.89}_{-4.32}$ & $68.28^{+75.52}_{-47.58}$ \\ 
        18.81 & 21.15 & Band+BB & 2710.11 & $-1.06^{+0.06}_{-0.07}$ & $-3.17^{+0.3}_{-0.34}$ & $98.89^{+5.99}_{-5.68}$ & $59.83^{+89.53}_{-38.74}$ \\ 
        21.15 & 22.38 & Band+BB & 1701.06 & $-0.94^{+0.13}_{-0.13}$ & $-2.92^{+0.27}_{-0.35}$ & $76.13^{+6.94}_{-5.78}$ & $79.11^{+72.9}_{-57.9}$ \\ 
        22.38 & 27.77 & Band+BB & 3957.98 & $-0.98^{+0.1}_{-0.09}$ & $-2.95^{+0.18}_{-0.25}$ & $52.58^{+2.38}_{-1.67}$ & $79.37^{+75.2}_{-60.53}$ \\ 
        27.77 & 30.91 & Band+BB & 3032.73 & $-1.3^{+0.11}_{-0.1}$ & $-3.35^{+0.29}_{-0.31}$ & $50.91^{+1.18}_{-0.67}$ & $74.96^{+80.0}_{-64.08}$ \\ 
        30.91 & 33.62 & Band+BB & 2714.41 & $-1.39^{+0.1}_{-0.07}$ & $-3.22^{+0.31}_{-0.35}$ & $51.38^{+1.7}_{-0.97}$ & $72.15^{+76.71}_{-56.63}$ \\ 
        33.62 & 37.96 & Band+BB & 3425.53 & $-1.45^{+0.07}_{-0.04}$ & $-2.87^{+0.34}_{-0.36}$ & $51.64^{+2.19}_{-1.2}$ & $81.73^{+73.91}_{-69.65}$ \\ 
        37.96 & 47.74 & Band+BB & 4643.79 & $-1.43^{+0.09}_{-0.05}$ & $-3.14^{+0.33}_{-0.36}$ & $51.77^{+2.63}_{-1.27}$ & $45.3^{+98.24}_{-38.87}$ \\ 
        47.74 & 58.90 & Band+BB & 4774.79 & $-1.36^{+0.17}_{-0.1}$ & $-2.9^{+0.41}_{-0.42}$ & $54.75^{+8.02}_{-3.57}$ & $43.25^{+93.44}_{-36.18}$ \\ 
\enddata
\tablecomments{Band: Band model; BB: blackbody; PL: power-law}
\end{deluxetable*}

\end{document}